\documentclass[prd,twocolumn,showpacs,amsmath,amssymb]{revtex4}
\usepackage{graphicx}
\usepackage{dcolumn}
\usepackage{bm}

\begin{document}

\title{Loop Quantum Cosmology of Bianchi I Model in $\bar{\mu}$ and $\bar{\mu}'$ Schemes with Higher Order Holonomy Corrections}
\author{Xiao-Jun Yue}
\email{yuexiaojun@mail.bnu.edu.cn}
 \affiliation{College of Information Engineering, Taiyuan University of Technology, Taiyuan 030024, China}
  \affiliation{Department of Physics, Beijing Normal University, Beijing 100875, China}
\author{Jian-Yang Zhu}
\thanks{Author to whom correspondence should be addressed}
\email{zhujy@bnu.edu.cn}
\affiliation{Department of Physics, Beijing Normal University, Beijing 100875, China}
\date{\today}

\begin{abstract}
The detailed formulation of loop quantum cosmology with higher order holonomy corrections has been constructed recently in the homogeneous and isotropic spacetime, yet it is important to extend the higher order holonomy corrections to include the effects of anisotropy which typically grow during the collapsing phase. In this paper we investigate the Bianchi I model in $\bar{\mu}'$ scheme which truly captures the regularization of the Hamiltonian constraint. To compare with the earlier works and provide a comparison with the $\bar{\mu}'$ scheme, we also investigate the $\bar{\mu}$ scheme although it has many disadvantages.   First we construct the effective dynamics with higher order holonomy corrections in a massless scalar field, then we extend it to the inclusion of arbitrary matter.  Besides that, we also analyze the behavior of the anisotropy during the evolution of the universe. We find that in the $\bar{\mu}'$ scheme,  the singularity is never approached and the quantum bounce is generic as in the isotropic case, regardless of the order of the holonomy corrections. Some differences in the bouncing phase of the two schemes are also found out.  It is also shown that in the two schemes the behavior of the anisotropy is not the same before and after the bounce.
\end{abstract}


\maketitle


\section{\label{s1}Introduction}

loop quantum gravity (LQG) is a mathematically
well-defined, nonperturbative, and background independent quantization of
gravity \cite{LQG}. The applications of LQG
to homogeneous and isotropic spacetime results in loop quantum cosmology (LQC). The
comprehensive formulation of LQC is constructed in the spatially flat,
isotropic model in detail \cite{LQC,LQC1,LQC2}, which indicates that the classical
big-bang singularity can be replaced by a big-bounce. With these successes, the methods can also be extended to Bianchi I model to
include anisotropy \cite{Chiou1,Ashtekar,Chiou2,Chiou3,Chiou4}.

The underlying dynamics in LQC is governed by the discreteness of the
quantum geometry. With the achievements described above, the scheme of LQC is indeed attractive, while its rigorous quantum approach
is difficult to afford due to the complexity. However, using semiclassical strategies we can
construct an effective method which has captured to a very good approximation in
the quantum dynamics \cite{Singh1}. With quantum corrections to the classical
Hamiltonian, we can get the effective equations of the modified Hamiltonian
which is an efficient approach to investigate the evolution of the early
universe. In \cite{Singh2} the evolution of the universe is investigated in
the form of effective approach, which indicates that the presence of a
big-bounce is a generic feature of LQC and does not require any exotic
matter that violates energy condition.

When the effective approach is extended to the anisotropic Bianchi I model,
the scheme is ambiguous in the early investigations as there are two strategies: $\bar{\mu}$ scheme and $%
\bar{\mu}^{\prime}$ scheme, each of which is of particular interest. In
\cite{Chiou2} and \cite{Chiou3} the $\bar{\mu}$ scheme is investigated and
in \cite{Chiou4,Singh,Singh01,Singh02,Singh03,Singh04} the $\bar{\mu}^{\prime}$ scheme is considered. The $\bar{\mu}$ scheme is just a simplifying assumption without a systematic justification\cite{Chiou1,Chiou3,Chiou4}. However,
this leads to drawbacks that the quantum dynamics depends on the choice of the fiducial cell\cite{Lukasz}. There are also problems that expansion scalar and shear scalar are not bounded above\cite{Singh}, which point towards lack of generic resolution of singularities in this quantization.  In \cite{Ashtekar} a more systematic procedure leads to the $\bar{\mu}'$
scheme, which is fiducial cell independent. This scheme also leads to the strong singularity resolution\cite{Singh,Singh01,Singh02}. Not only do
the results at the level of effective dynamics agree with the anticipations
of the rigorous quantum approach but some details of quantum effects during
the evolution are also obtained.

Despite the fascinating and attractive features, whether the quantum effects
result from the discreteness of the spacetime geometry of LQC is still
questionable, as some of the results through the rigorous quantum approach can
also be obtained through the heuristic effective dynamics in continuum
spacetime. In response to the deficiency, a new avenue of higher order
holonomy corrections is investigated \cite{Chiou5}.

The approach of the higher order holonomy corrections is also a heuristic
effective strategy that is more general than the conventional scheme, which
takes the traditional one to be the situation where the order of the
holonomy corrections is $0$. In \cite{Chiou5} it reveals that the
big-bounce are a generic feature of LQC no matter whether the
higher order holonomy corrections are included and the matter density remains finite with
an upper bound. In \cite{Chiou6} the rigorous quantum theory of LQC with
higher order holonomy corrections is formulated and the anticipations of
\cite{Chiou5} are confirmed. It is also shown that the higher order holonomy
corrections can be interpreted as a result of admitting generic $SU(2)$
representations for the Hamiltonian constraint operators.

The heuristic analysis of higher order holonomy corrections is a very
promising approach. However, the extension to the anisotropic case is still
an open issue. In this paper we investigate the higher order holonomy
corrections in Bianchi I model.

Firstly we construct the effective dynamics of Bianchi I model with higher
order holonomy corrections. We extend the $\bar{\mu}$ scheme to the
case of higher order holonomy corrections to compare with the early investigations and provide a  comparison with the $\bar{\mu}'$ scheme. On the other hand, although the numerical simulations about the bouncing phase of $\bar{\mu}'$ scheme are abundant, the analytical investigations about the this phase are some limited due to the mathematical complexity, especially with the arbitrary matter case.
 In this paper, we investigate the effective dynamics with arbitrary matter analytically in detail and extend it to
higher order holonomy corrections.

The anisotropy is also an important aspect as it grows in the contracting
phase of the evolution. Besides that, it can also tell us some information
of the universe before the big bounce. In this paper we also investigate the
evolution of the anisotropy with higher order holonomy corrections and
compare the differences of the behavior in the two schemes.

This paper is organized as follows. Firstly, we review briefly the classical dynamics in Bianchi I
model in Sec. \ref{s2}, and then, in Sec. \ref{s3}, we introduce the effective loop quantum dynamics with higher order holonomy
corrections and two sets of research schemes: $\bar{\mu}$ scheme and $\bar{\mu}^{\prime}$
scheme. Next in Sec. \ref{s4} and  Sec. \ref{s5}, we investigate in detail the
effective dynamics in the forms of $\bar{\mu}$ scheme and $\bar{\mu}^{\prime}$ scheme, respectively. In Sec. \ref{s6},
the anisotropies of the Bianchi I model in the two schemes are analyzed. Finally, we draw the conclusions
in Sec. \ref{s7}.

\section{\label{s2}Classical Dynamics}

In this section, we review briefly the classical dynamics in Bianchi I
model. As a comparison of previous investigations,we first focus on the
model with a massless scalar field. Then we will turn to the case where a
general matter potential is considered. For a more complete description of the classical dynamics one can see,e.g.\cite{Chiou2,Chiou3,Chiou4}.

The spacetime metric of Bianchi I model is given as
\begin{equation}
ds^2=-N^2dt^2+a_1^2dx^2+a_2^2dy^2+a_3^2dz^2,
\end{equation}
where $N$ is the lapse function. When we write the classical dynamics in the
Ashtekar variables, we consider the spacetime with a manifold $\Sigma \times %
\mathbb{R}$ where the space supersurface $\Sigma $ is flat. Because of the
non-compactness of the spatial manifold, it is necessary to introduce a
fiducial cell ${\cal V}$ which has a fiducial volume $V_0=l_1l_2l_3$. In
Bianchi I model, the Ashtekar variables take a simple form where the phase
space is given by the diagonal triad variables $p_I$ and diagonal connection
variables $c_I$ ($I=1,2,3$). The canonical conjugate phase space satisfies
\begin{equation}
\left\{ c_I,p_J\right\} =\kappa \gamma \delta _{IJ},
\label{canonicalrelation}
\end{equation}
where $\kappa =8\pi G$ and $\gamma$ is the Barbero-Immirzi
parameter which was set to be $\gamma \simeq 0.2375$ by the black hole thermodynamics \cite{BarberoImmirzi}. The triad $p_I$ are related to the scale factors $a_I$ by
\begin{equation}
\left| p_1\right| =l_2l_3a_2a_3,\left| p_2\right| =l_1l_3a_1a_3,\left|\label{p}
p_3\right| =l_1l_2a_1a_2.
\end{equation}
Thus the triad variables are the physical areas of the rectangular surface
of ${\cal V}$ which is invariant under the coordinate rescaling. The
connection variables are given by
\begin{equation}
c_1=\gamma l_1\dot{a}_1,c_2=\gamma l_2\dot{a}_2,c_3=\gamma l_3\dot{a}_3,
\label{classicalc}
\end{equation}
that is the time change rates of the physical lengths of the edges of ${\cal %
V}$, which is also invariant under the coordinate rescaling \cite{Chiou4}.
Thus, the Hamiltonian constraint in the Ashtekar variables can be written as
\begin{equation}
{\cal H}_{cl}=-\frac N{\kappa \gamma ^2V}\left(
c_1p_1c_2p_2+c_2p_2c_3p_3+c_3p_3c_1p_1\right) +{\cal H}_{matt},\nonumber
\end{equation}
where $V=l_1l_2l_3a_1a_2a_3$ is the physical volume of the fiducial cell $\mathcal{V}$ and ${\cal H}_{matt}$ is the matter Hamiltonian. The form of the matter
Hamiltonian is
\begin{equation}
{\cal H}_{matt}=N\sqrt{p_1p_2p_3}\rho _M.\nonumber
\end{equation}
Equations of motion are
\begin{equation}
\dot{p}_I =-\kappa \gamma \frac{\partial {\cal H}_{cl}}{\partial c_I},~~
\dot{c}_I =\kappa \gamma \frac{\partial {\cal H}_{cl}}{\partial p_I}.
\end{equation}

\subsection{\label{s2a}for a massless scalar field}

In order to compare with \cite{Chiou1,Chiou2} and \cite{Chiou5}, here we
focus on a massless scalar field. For simplicity we choose the lapse
function $N=\sqrt{p_1p_2p_3}$ and introduce a new time variable
 $dt'=(p_1p_2p_3)^{-1/2}dt$. The rescaled Hamiltonian is
\begin{equation}
{\cal H}_{cl}=-\frac 1{\kappa \gamma ^2}\left(
c_1p_1c_2p_2+c_2p_2c_3p_3+c_3p_3c_1p_1\right) +\frac{p_\phi ^2}2.\nonumber
\end{equation}
The equations of motion are:
\begin{equation}
\frac{d\phi }{dt^{\prime }}=p_\phi ,\quad \frac{dp_\phi }{dt^{\prime }}=0,
\label{clmassless1}
\end{equation}
\begin{equation}
\frac{dc_1}{dt^{\prime }}=-\gamma ^{-1}c_1(c_2p_2+c_3p_3),
\label{clmassless2}
\end{equation}
\begin{equation}
\frac{dp_1}{dt^{\prime }}=\gamma ^{-1}p_1(c_2p_2+c_3p_3).
\label{clmassless3}
\end{equation}
With Eq.(\ref{clmassless2}) and Eq.(\ref{clmassless3}) we have
\begin{equation}
\frac d{dt^{\prime }}\left( c_Ip_I\right) =0\Rightarrow c_Ip_I=\kappa \gamma
\hbar {\cal K}_I,  \label{clmassless4}
\end{equation}
where ${\cal K}_I$ are constants.  Here we define
\begin{equation}
p_\phi =\hbar \sqrt{\kappa }{\cal K}_\phi.\label{kphi}
\end{equation}
Using Eq.(\ref{clmassless4}), Eq.(\ref{kphi}) and the
Hamiltonian constraint ${\cal H}_{cl}=0$ we have
\begin{equation}
{\cal K}_\phi ^2=2({\cal K}_2{\cal K}_3+{\cal K}_3{\cal K}_1+{\cal K}_1{\cal %
K}_2).  \label{clmassless5}
\end{equation}
Combining Eq.(\ref{clmassless4}) and Eq.(\ref{clmassless3}) gives
\begin{equation}
\frac 1{p_1}\frac{dp_1}{dt^{\prime }}=\kappa \hbar \left( {\cal K}_2+{\cal K}%
_3\right) .
\end{equation}
With Eq.(\ref{clmassless1}) the above equation can be written to be
\begin{equation}
\frac 1{p_1}\frac{dp_1}{d\phi }=\kappa \hbar \frac{{\cal K}_2+{\cal K}_3}{%
p_\phi }=\sqrt{8\pi G}\left( \frac{1-\kappa _1}{\kappa _\phi }\right) ,
\end{equation}
where ${\cal K}_I={\cal K}\kappa _I$, ${\cal K}_\phi ={\cal K}\kappa _\phi $
and
\begin{equation}
\kappa _1+\kappa _2+\kappa _3=1,\kappa _1^2+\kappa _2^2+\kappa _3^2+\kappa
_\phi ^2=1.  \label{clmassless6}
\end{equation}

\subsection{\label{s2b}for an arbitrary matter field}

Now we consider the inclusion of an arbitrary matter field. Here we also choose the
lapse function $N=\sqrt{p_1p_2p_3}$ and the Hamiltonian takes the form
\begin{equation}
{\cal H}_{cl}=-\frac 1{\kappa \gamma ^2}\left(
c_1p_1c_2p_2+c_2p_2c_3p_3+c_3p_3c_1p_1\right) +p_1p_2p_3\rho _M.\nonumber
\end{equation}
The equations of motion are:
\begin{equation}
\frac{dp_1}{dt^{\prime }}=\gamma ^{-1}p_1\left( c_2p_2+c_3p_3\right) ,
\label{clarbitrary1}
\end{equation}
\begin{eqnarray}
\frac{dc_1}{dt^{\prime }} &=&-\gamma ^{-1}c_1\left( c_2p_2+c_3p_3\right)
\nonumber \\
&&+\kappa \gamma p_2p_3\left( \rho _M+p_1\frac{\partial \rho _M}{\partial p_1%
}\right) .  \label{clarbitrary2}
\end{eqnarray}
One can see that when the energy density $\rho _M=p_\phi ^2/2(p_1p_2p_3)$ Eq.(\ref
{clarbitrary2}) turns out to be Eq.(\ref{clmassless2}). Combining Eq.(\ref
{clarbitrary1}) and Eq.(\ref{clarbitrary2}) gives the relation
\begin{equation}
\frac d{dt^{\prime }}(p_Ic_I)=\kappa \gamma p_1p_2p_3\left(\rho _M+p_I\frac{%
\partial \rho _M}{\partial p_I}\right).
\end{equation}
If we assume that the matter has zero anisotropy, namely $%
\rho _M(p_1,p_2,p_3)=\rho _M(p_1p_2p_3)$, we can get $p_I\frac{\partial
\rho _M}{\partial p_I}=p_J\frac{\partial \rho _M}{\partial p_J}$, which
yields
\begin{equation}
\frac d{dt^{\prime }}\left( p_Ic_I-p_Jc_J\right) =0.  \label{clarbitrary3}
\end{equation}
The above equation can be integrated to be:
\begin{equation}
p_Ic_I-p_Jc_J=\gamma V\left( H_I-H_J\right) =\gamma V_0\alpha _{IJ}.\label{clarbitrary4}
\end{equation}
where $\alpha _{IJ}$ is a constant anti-symmetric matrix. From the
constraint ${\cal H}_{cl}=0$ we can get the relation
\begin{equation}
H_1H_2+H_2H_3+H_3H_1=\kappa \rho _M.\nonumber
\end{equation}
We can also define the mean scale factor $a$ as
\begin{equation}
a=\left( a_1a_2a_3\right) ^{1/3},\label{a}
\end{equation}
then
\begin{equation}
H=\frac{\dot{a}}a=\frac 13\left( H_1+H_2+H_3\right)
\end{equation}
is the mean Hubble rate. In some literatures the expansion scalar is also defined to be $\theta=\frac{1}{V}\frac{dV}{dt}=3H$ to describe the expansion rate of the whole volume.

 The Friedmann equation with the inclusion of anisotropy is
\begin{eqnarray}
H^2&&=\frac 13\left( H_1H_2+H_2H_3+H_3H_1\right)   \nonumber \\
&&+\frac 1{18}\left[ \left( H_1-H_2\right) ^2+\left( H_2-H_3\right)
^2+\left( H_3-H_1\right) ^2\right]    \nonumber \\
&&=\frac \kappa 3\rho _M+\frac{\Sigma ^2}{a^6},  \label{clFriedmann}
\end{eqnarray}
 The shear parameter is
\begin{equation}
\Sigma ^2=\frac 1{18}\left( \alpha _{12}^2+\alpha _{23}^2+\alpha
_{31}^2\right) ,  \label{clshear}
\end{equation}
which is a constant in the classical case. The anisotropic shear scalar $%
\sigma ^2=\sigma _{\mu \nu }\sigma ^{\mu \nu }$ is given by
\begin{eqnarray}
\sigma ^2 &=&\frac 13\left[ \left( H_1-H_2\right) ^2+\left( H_2-H_3\right)
^2+\left( H_3-H_1\right) ^2\right]   \nonumber \\
&=&\frac{6\Sigma ^2}{a^6}.  \label{shearscalar}
\end{eqnarray}

\section{\label{s3}Effective Loop quantum dynamics}

In the effective dynamics of LQC, the connection variables $c_I$ ($I=1,2,3$) should be replaced by
holonomies, i.e.,
\begin{equation}
c_I\rightarrow \frac{\sin \left( \bar{\mu}_Ic_I\right) }{\bar{\mu}_I},
\end{equation}
where $\bar{\mu}_I$ are real valued functions of $p_I$ which are measures
of the discreteness in the quantum gravity. When $\bar{\mu}_I\ll 1$, $\sin (%
\bar{\mu}_Ic_I)/\bar{\mu}_I\approx c_I$.

By choosing this, the Hamiltonian can be written as
\begin{eqnarray}
{\cal H} &=&-\frac N{\kappa \gamma ^2V}\left[ \frac{\sin \left( \bar{\mu}%
_1c_1\right) }{\bar{\mu}_1}\frac{\sin \left( \bar{\mu}_2c_2\right) }{\bar{\mu%
}_2}p_1p_2\right.   \nonumber \\
&&+\frac{\sin \left( \bar{\mu}_2c_2\right) }{\bar{\mu}_2}\frac{\sin \left(
\bar{\mu}_3c_3\right) }{\bar{\mu}_3}p_2p_3  \nonumber \\
&&\left. +\frac{\sin \left( \bar{\mu}_3c_3\right) }{\bar{\mu}_3}\frac{\sin
\left( \bar{\mu}_1c_1\right) }{\bar{\mu}_1}p_3p_1\right] +{\cal H}_{matt}.\label{conHamiltonian}
\end{eqnarray}
In this paper we consider the effective dynamics with higher order holonomy
corrections. In fact, it is possible to approximate $c_I$ in terms of $\sin (%
\bar{\mu}_Ic_I)$ to arbitrary accuracy
\begin{equation}
c_I=\frac 1{\bar{\mu}_I}\sum_{k=0}^\infty \frac{\left( 2k\right) !}{%
2^{2k}\left( k!\right) ^2\left( 2k+1\right) }\left[ \sin \left( \bar{\mu}%
_Ic_I\right) \right] ^{2k+1}.\label{c_I}
\end{equation}
This inspires us to define the $n$th order holonomized connection variables\emph{}
as
\begin{equation}
c_I^{(n)}:=\frac 1{\bar{\mu}_I}\sum_{k=0}^n\frac{\left( 2k\right) !}{%
2^{2k}\left( k!\right) ^2\left( 2k+1\right) }\left[ \sin (\bar{\mu}%
_Ic_I)\right] ^{2k+1},  \label{higherholonomy}
\end{equation}
which  remains a periodic and
bounded function of $c_I$. The conventional holonomy correction corresponds to $n=0$.

The remarkable point is that in fact it is only when $-\pi/2\leqslant \bar{\mu}_Ic_I\leqslant\pi/2$ does the power series give back to $c_I$. As $c_I^{(n)}$ is a periodic function of $c_I$, when $\bar{\mu}_Ic_I$ exceeds this regime, the $c_I^{(n)}$ does not blow up as $c_I$, it is still bounded, even if $n\rightarrow \infty$. In the following we will see that it is the $\bar{\mu}_Ic_I$ goes through the point $\pi/2$ that makes the cosine function flip its sign, which leads to the big bounce. This interprets the reason that when $n\rightarrow \infty$ the dynamics does not reduce to the classical case, as after the bouncing point $\bar{\mu}_Ic_I>\pi/2$ the variable $c_I^{(\infty)}$ does not reduce to $c_I$,  the classical dynamics is still modified drastically in the limit $n\rightarrow\infty$.

With this, the Hamiltonian with holonomy corrections up to the $n$th order
can be written as
\begin{eqnarray}
{\cal H}_{eff} &=&-\frac N{\kappa \gamma ^2V}\left[
c_1^{(n)}p_1c_2^{(n)}p_2+c_2^{(n)}p_2c_3^{(n)}p_3\right.   \nonumber \\
&&\left. +c_3^{(n)}p_3c_1^{(n)}p_1\right] +{\cal H}_{matt}.
\label{Hamiltonian}
\end{eqnarray}

Using the canonical relation (\ref{canonicalrelation}) we can get the
relation
\begin{equation}
\left\{ c_I,c_J^{(n)}\right\} =\frac{\kappa \gamma }{\bar{\mu}_J}\frac{%
\partial \bar{\mu}_J}{\partial p_I}\left[ \cos \left( \bar{\mu}_Jc_J\right) %
\mathfrak{S}_n\left( \bar{\mu}_Jc_J\right) c_J-c_J^{(n)}\right]
\label{higherordercanonicalrelation}
\end{equation}
\begin{equation}
\left\{ p_I,c_J^{(n)}\right\} =-\kappa \gamma \cos (\bar{\mu}_Jc_J)%
\mathfrak{S}_n(\bar{\mu}_Jc_J)\delta _{IJ},
\end{equation}
where
\begin{eqnarray}
\mathfrak{S}_n\left( \bar{\mu}_Ic_I\right)  &:&=\sum_{k=0}^n\frac{(2k)!}{%
2^{2k}\left( k!\right) ^2}\sin (\bar{\mu}_Ic_I)^{2k}  \nonumber \\
&&\stackrel{n\rightarrow \infty }{\rightarrow }\left| \cos (\bar{\mu}%
_Ic_I)\right| ^{-1}.
\end{eqnarray}

\subsection{\label{s3a}The meaning of higher order holonomy corrections}
The initial motivation of introducing higher order holonomy corrections is to provide further evidence that the quantum bounce is a consequence of the intrinsic discreteness of geometry in LQC\cite{Chiou5}, but the  implication is not just limited to the heuristic effective level. In \cite{Chiou6} the rigorous quantum theory with higher order holonomy corrections is formulated in the isotropic model. Until now the quantum theory of higher order holonomy corrections in Bianchi I model is not completed, and this construction is not the focus of this paper, but taking a look at the quantum theory  in the isotopic model can still provide us a taste of the meaning of the higher order holonomy corrections. The detailed analysis can be found in \cite{Chiou6}.

It was shown that the higher order holonomy  corrections is related to the $j$ ambiguity. We know that in LQC the classical variable $c$ should be replaced by the holonomy
\begin{equation}
h_l^{(\bar{\mu})}:=\mathcal{P}exp\int^{\bar{\mu}L}_0\tau_i A_a^i dx^a=exp(\bar{\mu}c\tau_i),\nonumber
\end{equation}
which is the holonomy along the edge of coordinate length $\bar{\mu}L$. Here $\tau_i$ are the $SU(2)$ generators with $[\tau_i,\tau_j]=\epsilon_{ij}^k\tau_k$. In standard convention $2i\tau_i=\sigma_i$ are the Pauli matrices. In fact,  the Lie algebra generators $\tau_i$ in the $j$ representation of the $SU(2)$ group are represented as $(2j+1)\times (2j+1)$ matrices $^{(j)}\tau_i$, which satisfy
\begin{equation}
Tr(^{(j)}\tau_i^{(j)}\tau_j)=-\frac{1}{3}j(j+1)(2j+1)\delta_{ij}\nonumber
\end{equation}
where $j$ is a half integer and the conventional choice is  $j=1/2$. The holonomies $h_i^{(\bar{\mu})}$
 can also be promoted to $^{(j)}h_i^{(\bar{\mu})}$ in the $j$ representation. We write $^{(j)}\widehat{C}'_{grav}$ to be the Hamiltonian operator in generic $j$ representations, which consists of even powers of $\widehat{\sin(\bar{\mu}c)}$. When $j=1/2$,
 \begin{equation}
 ^{(1/2)}\widehat{C}'_{grav}=-\frac{3}{\kappa\gamma^2\bar{\mu}^2}\widehat{\sin(\bar{\mu}c)}^2|\widehat{p}|^2\nonumber
 \end{equation}
which leads to the conventional holonomy corrections ($c^{(n=0)}$).

In \cite{Chiou6} it was proved that the gravitational part of Hamiltonian operator with higher order holonomy corrections $\hat{C}_{grav}^{'(n)}$ can be represented as a linear superposition of the Hamiltonian operators in generic $j$ representations:
\begin{equation}
\widehat{C}^{'(n)}_{grav}=\sum_{j=1/2}^{n/2}c_j^{(n)(j)}\widehat{C}'_{grav},\nonumber
\end{equation}
where $c_j^{(n)}$ are constants.  In the limit $n\rightarrow\infty$, we have to include all the $j$ representations.

There is a problem that although the linear superposition coefficients $c_j^{(n)}$  are constants for a definite $n$,
they diverge as $n\rightarrow\infty$. However, the computed data has shown that comparatively only the contributions from the moderate $j$ are appreciable. If we compose a regularization to supress the high $j$ contributions, the coefficients
$c_j^{(n)}$ can be made to converge. With a suitable regularization, the $\widehat{C}^{'(\infty)}_{grav}$ can be obtained.

In \cite{Chiou6} the relation between higher order holonomy corrections and spin-foams is discussed briefly. The peculiar feature of the divergence of $c_j^{(n)}$ is reminiscent to the infrared divergence in many spin-foam models\cite{Ponzano,Freidel,Baez}, and the regularization may correspond to a nonzero cosmological constant\cite{Noui}. It's reasonable to speculate that the higher order holonomy corrections comes out from the spin-foam models more naturally. The research about this issue may help us to find out the link between LQC and the spin-foam formalism.

In \cite{Vandersloot} and \cite{Perez} it was shown that the  $j>1/2$ representations will lead to the ill-behaving spurious solutions. However, the investigation in \cite{Chiou6} has shown that the expectation values of Dirac observables are well behaved when we consider $\widehat{C}^{'(n)}_{grav}$, at least for the case  $n=\infty$. The reason may be that the spurious solutions come out only in a definite $j$ representation and suppressed when all $j$ representations are included to match the expression $\widehat{C}^{'(n)}_{grav}$. There are also signs which bolster that it may be more natural to include all $j$ representations, although no theory suggests it in first principle.

The detailed quantum theory in Bianchi I model with $j=1/2$ is also formulated in \cite{Ashtekar}, but the construction of the quantum theory with higher order holonomy corrections in Bianchi I model is still an open issue. It is very probable that in the Bianchi I model the higher order holonomy corrections can also be interpreted as a result of admitting generic $SU(2)$ representations for the Hamiltonian operator.

 In \cite{Ashtekar} a  semiheuristic consideration of the well-motivated correspondence between kinematic states in LQG and LQC suggests that $j=1/2$. The reason is that although the macroscopic geometry is spatially homogeneous, the microscopic is not exactly homogeneous. To achieve the best possible coarse that constitute the large scale geometry, the edges should be packed as tightly as possible. However, if we focus our attention on the precisely homogeneous Bianchi I model, the former requirement seems unnecessary. This is the question that whether the investigations  for the theory of LQC are derived from LQG or within the confines of LQC. As a  systematic formulation to derive LQC from LQG is still an open issue, it is logically legitimate to reduce the theory to homogeneous level first and then quantize.

The implications of the higher order holonomy corrections in the quantum level is still not conclusive, the linear superposition of generic $j$ representations is just one possibility.  This issue still requires future researches, which is not the focus of this paper. In the following we only focus upon the effective dynamics.

\subsection{\label{s3b}The $\bar{\mu}$ and $\bar{\mu}'$ schemes}
We can see from Eq. (\ref{higherordercanonicalrelation}) that the
equation is different for different expressions of $\bar{\mu}_I$. In the
isotropic case, the discreteness variable  $\bar{\mu}$ has the form  $\bar{\mu}\propto 1/\sqrt{p}$.

In Bianchi I model, the schemes are more ambiguous. When the anisotropy is considered, there are three $p_I$ and three different $\bar{\mu}_I$ should be introduced. Generally speaking there are two schemes which can reduce to the the consistent isotropic case:
\begin{itemize}
\item  {$\bar{\mu}$ scheme:}
\begin{equation}
\bar{\mu}_1=\sqrt{\frac{\triangle }{p_1}},\bar{\mu}_2=\sqrt{\frac{\triangle
}{p_2}},\bar{\mu}_3=\sqrt{\frac{\triangle }{p_3}}
\end{equation}
\item  {$\bar{\mu}^{\prime }$ scheme:}
\begin{equation}
\bar{\mu}_1^{\prime }=\sqrt{\frac{\triangle p_1}{p_2p_3}},\bar{\mu}%
_2^{\prime }=\sqrt{\frac{\triangle p_2}{p_3p_1}},\bar{\mu}_3^{\prime }=\sqrt{%
\frac{\triangle p_3}{p_1p_2}}.
\end{equation}
\end{itemize}
Here $\triangle$ is the area gap in LQG. However, as will be seen the $\bar{\mu}$ scheme  has the drawbacks that its effective
dynamics is dependent of the choice of $\mathcal{V}$\cite{Lukasz}. In \cite{Ashtekar} the $\bar{\mu}'$
scheme was introduced by a more systematic procedure, which is fiducial cell independent. This can be easily seen from the simplest form of Hamiltonian, Eq.(\ref{conHamiltonian}), which corresponds to $H_{eff}^{(n=0)}$. In this form, the $\sin(\bar{\mu}_Ic_I)$ is actually related to the shift operator in the full quantum theory. When $\bar{\mu}_Ic_I$ changes, the volume shift corresponding to the shift operator varies. Considering $\bar{\mu}$ scheme, we have $\bar{\mu}_1c_1=\frac{\sqrt{\triangle}c_1}{p_1^{1/2}}$. If we choose a different fiducial volume
\begin{equation}
V_0=l_1l_2l_3\rightarrow V_0'=L_1l_1L_2l_2L_3l_3=L_1L_2L_3V_0,\nonumber
\end{equation}
 it can be seen from Eq.(\ref{p}) and Eq.(\ref{classicalc}) that
\begin{equation}
\bar{\mu}_1c_1\rightarrow\bar{\mu}_1c_1\frac{L_1}{\sqrt{L_2L_3}}\nonumber
\end{equation}
which changes its value. On the other hand, when we consider $\bar{\mu}'$ scheme with $\bar{\mu}'_1c_1=\frac{\sqrt{\triangle}p_1^{1/2}c_1}{(p_2p_3)^{1/2}}$, one can verify that  this problem disappears.

In addition to the problem of rescaling under shape of the fiducial cell, the $\bar{\mu}$ scheme has other problems, including the expansion and shear scalar are not bounded above as documented in\cite{Singh}. This means that even when there might be solutions of 'bounce' to the effective equations of motion, there is no universal 'quantum gravity scale', which leads to the lack of resolution of the strong singularity. Nowadays all kinds of signs indicate that the $\bar{\mu}'$ scheme is much more preferable.

\section{\label{s4}Effective dynamics in $\bar{\mu}$ Scheme}

In this section, we will construct the effective dynamics in $\bar{\mu}$
scheme with higher order holonomy corrections in a massless scalar field and
in an arbitrary scalar field respectively.

In $\bar{\mu}$ scheme, Eq.(\ref{higherordercanonicalrelation}) turns out to
be
\begin{eqnarray}
\left\{ c_I,c_J^{(n)}\right\} &=&-\kappa \gamma \frac 1{%
2p_J}\left[ \cos \left( \bar{\mu}_Jc_J\right) \mathfrak{S}_n\left( \bar{\mu}%
_Jc_J\right) c_J-c_J^{(n)}\right] \delta _{IJ}.  \nonumber \\
&&~
\end{eqnarray}

In the effective dynamics we also choose the lapse function to be $N=\sqrt{%
p_1p_2p_3}$ and Eq.(\ref{Hamiltonian}) becomes
\begin{eqnarray}
{\cal H}_{\bar{\mu}} &=&-\frac 1{\kappa \gamma ^2}\left[
c_2^{(n)}p_2c_3^{(n)}p_3+c_3^{(n)}p_3c_1^{(n)}p_1\right.   \nonumber \\
&&\left. +c_1^{(n)}p_1c_2^{(n)}p_2\right] +p_1p_2p_3\rho _M.
\label{barHamiltonian}
\end{eqnarray}
The equations of motion are
\begin{equation}
\frac{dp_1}{dt^{\prime }}=\frac 1\gamma \cos \left( \bar{\mu}_1c_1\right) %
\mathfrak{S}_n\left( \bar{\mu}_1c_1\right) p_1\left[
c_3^{(n)}p_3+c_2^{(n)}p_2\right] ,  \label{bar1}
\end{equation}
\begin{eqnarray}
\frac{dc_1}{dt^{\prime }} &=&-\frac 1\gamma \left[ \frac 32c_1^{(n)}-\frac 12%
\cos \left( \bar{\mu}_1c_1\right) \mathfrak{S}_n\left( \bar{\mu}_1c_1\right)
c_1\right]   \nonumber \\
&&\times \left[ p_2c_2^{(n)}+p_3c_3^{(n)}\right] +\kappa \gamma p_2p_3\left[
\rho _M+p_1\frac{\partial \rho _M}{\partial p_1}\right] .  \nonumber\\
~\label{bar2}
\end{eqnarray}
We also have
\begin{eqnarray}
\frac{dc_1^{(n)}}{dt^{\prime }} &=&-\frac 1\gamma c_1^{(n)}\cos \left( \bar{%
\mu}_1c_1\right) \mathfrak{S}_n\left( \bar{\mu}_1c_1\right) (c_2^{(n)}p_2+c_3^{(n)}p_3)
\nonumber \\
&&+\kappa \gamma \cos(\bar{\mu}_1c_1)\mathfrak{S}_n\left( \bar{\mu}_1c_1\right) p_2p_3\left(
\rho _M+p_1\frac{\partial \rho _M}{\partial p_1}\right) . \nonumber\\
~ \label{bar3}
\end{eqnarray}
With Eq.(\ref{bar1}) and Eq.(\ref{bar3}) we can get
\begin{eqnarray}
\frac d{dt^{\prime }}\left( p_1c_1^{(n)}\right)  &=&\kappa \gamma %
\mathfrak{S}_n(\bar{\mu}_1c_1)\cos (\bar{\mu}_1c_1)  \nonumber \\
&&\times p_1p_2p_3\left( \rho _M+p_1\frac{\partial \rho _M}{\partial p_1}%
\right) .  \label{bar4}
\end{eqnarray}

\subsection{\label{s4a}for a massless scalar field}

In the case of a massless scalar field, the energy density is $\rho _M=\frac{p_\phi ^2}{2p_1p_2p_3}$. We can get
\begin{equation}
\frac{dp_\phi }{dt^{\prime }}=\left\{ p_\phi ,{\cal H}_{eff}\right\} =0,
\label{barmassless1}
\end{equation}
and
\begin{equation}
\frac{d\phi }{dt^{\prime }}=\left\{ \phi ,{\cal H}_{eff}\right\} =p_\phi .
\label{barmassless2}
\end{equation}
Eq.(\ref{barmassless1}) means that $p_\phi $ is a constant and Eq.(\ref
{barmassless2}) shows that $\phi $ can be regarded as an emergent time. With a
massless scalar field Eq.(\ref{bar4}) has the form
\begin{equation}
\frac d{dt^{\prime }}\left( p_Ic_I^{(n)}\right) =0,  \label{barmassless3}
\end{equation}
which means
\begin{equation}
p_Ic_I^{(n)}=\kappa \gamma \hbar {\cal K}_I  \label{barmassless4}
\end{equation}
where ${\cal K}_I$ is a constant. Comparing Eq.(\ref{barmassless4}) with Eq.(%
\ref{clmassless4}) it can be seen that the classical connection $c_I$ is
replaced by $c_I^{(n)}$. From the constraint ${\cal H}_{eff}=0$ we can
get the relation
\begin{equation}
p_\phi ^2=\frac 2{\kappa \gamma ^2}\left[
c_1^{(n)}p_1c_2^{(n)}p_2+c_2^{(n)}p_2c_3^{(n)}p_3+c_3^{(n)}p_3c_1^{(n)}p_1%
\right] .
\end{equation}
As in the classical case, by defining $p_\phi =\hbar \sqrt{\kappa }{\cal K}_\phi $, we can get Eq.(\ref{clmassless5}) and Eq.(\ref{clmassless6}).
Combining Eq. (\ref{bar1}) and Eq. (\ref{barmassless4}) yields
\begin{equation}
\frac 1{p_1}\frac{dp_1}{dt^{\prime }}=\kappa \hbar \cos (\bar{\mu}_1c_1)%
\mathfrak{S}_n(\bar{\mu}_1c_1)\left( {\cal K}_2+{\cal K}_3\right) .
\end{equation}
By regarding $\phi $ as an emergent time, via Eq.(\ref{barmassless2}) we can
obtain
\begin{equation}
\frac 1{p_1}\frac{dp_1}{d\phi }=\sqrt{\kappa }\cos (\bar{\mu}_1c_1)%
\mathfrak{S}_n(\bar{\mu}_1c_1)\frac{1-\kappa _1}{\kappa _\phi }.\label{movingeq}
\end{equation}
When $\bar{\mu}_1c_1\ll 1$, $\cos (\bar{\mu}_1c_1)\rightarrow 1$, $\sin (%
\bar{\mu}_1c_1)\rightarrow 0$, $\mathfrak{S}_n\rightarrow 1$ and $%
c_I^{(n)}\rightarrow c_I$, Eq.(\ref{bar1}), Eq.(\ref{bar2}) and Eq.(\ref{bar3})
all turn out to be the classical form. On the other hand, when $\bar{\mu}%
_Ic_I$ is significant, the quantum corrections are more and more
appreciable. When $\cos (\bar{\mu}_1c_1)=0$ (i.e., $\bar{\mu}_1c_1=\pi /2$),
the quantum bounce occurs. From Eq.(\ref{higherholonomy}) this happens when
\begin{equation}
c_1^{(n)}\bar{\mu}_1=\sum_{k=0}^n\frac{(2k)!}{2^{2k}\left( k!\right) ^2(2k+1)%
}=:\mathfrak{F}_n,\label{Fn}
\end{equation}
where $\mathfrak{F}_n\rightarrow \pi /2$ as $n\rightarrow \infty $. From Eq.(%
\ref{barmassless4}) and Eq.(\ref{higherholonomy}), we can also get
\begin{eqnarray}
p_1^{3/2} &=&\sqrt{\triangle }\frac{\kappa \gamma \hbar {\cal K}_1}{%
\sum_{k=0}^n\frac{\left( 2k\right) !}{2^{2k}\left( k!\right) ^2(2k+1)}\left[
\sin \left( \bar{\mu}_1c_1\right) \right] ^{2k+1}}  \nonumber \\
&\geqslant &\sqrt{\triangle }\frac{\kappa \gamma \hbar {\cal K}_1}{%
\mathfrak{F}_n}.  \label{308}
\end{eqnarray}
In the second step the equality holds when $\bar{\mu}_1c_1=\pi /2$,
which is just the bounce point. One can see that the classical singularity
is never approached and the bounce is robust under the inclusion of
anisotropies. We can define the directional density:
\begin{equation}
{\cal \rho }_I:=\frac{p_\phi ^2}{p_I^3}
\end{equation}
for the $I$-direction and its critical value is
\begin{equation}
{\cal \rho }_{I,crit}:=\frac{p_\phi ^2}{p_{I,bounce}^3}=\mathfrak{F}%
_n^2\left( \frac{\kappa _\phi }{\kappa _I}\right) ^2\rho _{Pl},\label{mubarrhoc}
\end{equation}
where $\rho _{Pl}:=(\kappa \gamma ^2\triangle )^{-1}$. It shows that the
evolutions of $p_I$ are decoupled in three different directions. Thus the
bounces occur up to three times, whenever each of the directional density
reaches its critical density, which is the same as the conventional holonomy
corrected case (corresponding to $n=0$ in higher order holonomy corrections). But the critical value of
directional density is different from the previous case.

The mean scale factor $a(t)$ is depicted in Fig.\ref{fig01}(a). It demonstrates
that the nonsingular bouncing scenario is robust regardless of $n$. The
quantum bounce of $a(t)$ occurs more abruptly as $n$ increases, and if $%
n\rightarrow\infty$, the bounce takes place so abruptly that it only
imprints a kink, which does not reduce to the classical form. This can be seen from Eq.(\ref{movingeq}). When $n\rightarrow \infty$, the term $\mathfrak{S}\rightarrow|\cos(\bar{\mu}_Ic_I)|^{-1}$, and the product $\cos(\bar{\mu}_Ic_I)\mathfrak{S}(\mu_Ic_I)\rightarrow sgn(\cos(\bar{\mu}_Ic_I))$. When $\bar{\mu}_Ic_I=\pi/2$, the cosine function flips its sign and the big-bounce happens abruptly.

\subsection{\label{s4b}for an arbitrary matter field}

In this section we consider the dynamics with arbitrary matter. Here we use
the method provided in \cite{Chiou3}. We can assume that the matter density
is in the form
\begin{equation}
\rho _M=A\left( p_1p_2p_3\right) ^{-(1+w)/2},  \label{rho}
\end{equation}
with $A$ a constant and $w$ the state parameter. When $a\rightarrow \infty $, the derivation in \cite{Chiou3} has shown that the effective dynamics reduces to the classical form when $-1<w<1$. Although here we use the higher
order holonomy corrections, this conclusion is still correct.

Then we consider the other limit $a\rightarrow 0$. When we consider the arbitrary matter,
Eq.(\ref{barmassless3}) is not satisfied, but we can still write it to be a simple form. We know that in the classical limit, $c_I^{(n)}\rightarrow c_I$, and we have Eq.(\ref{clarbitrary3}), which leads to $p_Ic_I-p_Jc_J=\gamma V_0\alpha_{IJ}$ as Eq.(\ref{clarbitrary4}).
Now for convenience we can write the integration constant $\gamma V_0\alpha_{IJ}$ to be $\gamma V_0\alpha_{IJ}=\kappa \gamma \hbar(\mathcal{K}_I-\mathcal{K}_J)$, where $\mathcal{K}_I$ and $\mathcal{K}_J$ are constants. With this, in the semiclassical regime, we can assume that
\begin{equation}
p_Ic_I^{(n)}=\kappa \gamma \hbar \left[ {\cal K}_I+f_I(t)\right] .
\label{bararbitrary1}
\end{equation}
The first term is a constant introduced in the classical case and the second term is time-dependent.
From Eq.(\ref{bar4}) and the expression of $\rho _M$ we can get
\begin{eqnarray}
&&\frac d{dt^{\prime }}\left( p_Ic_I^{(n)}-p_Jc_J^{(n)}\right)  \nonumber \\
&=&\kappa \gamma \frac{1-w}2A\left[ \mathfrak{S}_n(\bar{\mu}_Ic_I)\cos (\bar{%
\mu}_Ic_I)\right.  \nonumber \\
&&\left. -\mathfrak{S}_n(\bar{\mu }_Jc_J)\cos (\bar{\mu }_Jc_J)\right]
\left( p_1p_2p_3\right) ^{\frac{1-w}2}.  \nonumber \\
&&~  \label{bararbitrary2}
\end{eqnarray}
One can see when $\bar{\mu}_Ic_I\rightarrow 0$ the above equation reduces to
Eq.(\ref{clarbitrary3}). When $a\rightarrow 0$, we also have $p_I\rightarrow
0$. If $w<1$, the above equation turns out to be $\frac d{dt^{\prime }}%
\left( p_Ic_I^{(n)}-p_Jc_J^{(n)}\right) \approx 0$, which means that $f_I(t)$
in Eq.(\ref{bararbitrary1}) has the same value near the bouncing point: $%
f_1(t)=f_2(t)=f_3(t)=f(t)$. The Hamiltonian constraint ${\cal H}_{eff}=0$
with ${\cal H}_{eff}$ given by Eq.(\ref{barHamiltonian}) then yields
\begin{eqnarray}
&&3f^2(t)+2\left( {\cal K}_1+{\cal K}_2+{\cal K}_3\right) f(t) \nonumber\\
&+&{\cal K}_2{\cal K}_3+{\cal K}_3{\cal K}_1+{\cal K}_1{\cal K}_2 =\frac A{\kappa \hbar ^2}\left( p_1p_2p_3\right) ^{\frac{1-w}2}.
\end{eqnarray}
The time-independent part satisfies
\begin{equation}
{\cal K}_2{\cal K}_3+{\cal K}_3{\cal K}_1+{\cal K}_1{\cal K}_2=0,
\label{bararbitrary3}
\end{equation}
and the time-dependent part is given by:
\begin{eqnarray}
f(t) &=&-\frac{{\cal K}_1+{\cal K}_2+{\cal K}_3}3  \nonumber \\
&&\pm \frac 13\left[ \left( {\cal K}_1+{\cal K}_2+{\cal K}_3\right) ^2+\frac{%
3A\left( p_1p_2p_3\right) ^{\frac{1-w}2}}{\kappa \hbar ^2}\right] ^{1/2}
\nonumber \\
&=&-\frac{{\cal K}}3\pm \frac 13\left[ {\cal K}^2+\frac{3A\left(
p_1p_2p_3\right) ^{\frac{1-w}2}}{\kappa \hbar ^2}\right] ^{1/2}.
\label{bararbitrary4}
\end{eqnarray}
In the second step we scale the constants ${\cal K}_I={\cal K}\kappa _I$ such that
Eq.(\ref{bararbitrary3}) gives
\begin{equation}
\kappa _1+\kappa _2+\kappa _3=1,\kappa _1^2+\kappa _2^2+\kappa _3^2=1.
\label{bararbitrary5}
\end{equation}
In Eq.(\ref{bararbitrary4}) we can choose the $+$ sign only without losing
any generality \cite{Chiou3}. When we discuss the bounces in the Bianchi I model,
we follow \cite{Chiou3} to separate three cases: (i) the Kasner phase, (ii)
the isotropized phase, and (iii) the transition phase.

The distinction between the Kasner phase, the isotropized phase and the transition phase is due to the classical dynamics in Bianchi I model. The detailed dynamics in classical case is investigated in Appendix A of \cite{Chiou3}. Here we just cite the conclusions in general terms. In the classical case of Bianchi I model with arbitrary matter,
the evolutions of  different $p_I$ are dominated by two parts: the anisotropic part $\mathcal{K}$ and the matter part $\rho_M$. If the matter part is negligible compared to the anisotropic part, the evolutions of $p_I$ are like the Kasner solution, which is called the Kasner phase. On the other hand, if the matter part is dominant, the variation rates of different $p_I$ are nearly the same, which is called the isotropized phase. The situation in between is called the transition phase. Here we have to note that even if in the isotropized phase, the universe is not exactly isotopic. It is just more isotropic than the Kasner phase.  Only if the initial conditions of the three directions are all the same does the Bianchi I model reduce to the isotropic model. In the following we will see that in LQC the case is the same.

In case (i)(the Kasner phase), the contribution from the matter sector is negligible and the
evolution is dominated by the constant ${\cal K}$. Thereby
\begin{eqnarray}
f(t) &\approx &-\frac{{\cal K}}3+\frac{{\cal K}}3\left[ 1+\frac{%
3A(p_1p_2p_3)^{\frac{1-w}2}}{2{\cal K}^2\kappa \hbar ^2}\right]  \nonumber \\
&\approx &\frac A{2\kappa {\cal K}\hbar ^2}(p_1p_2p_3)^{\frac{1-w}2}.
\label{bararbitrary6}
\end{eqnarray}
Applying it to Eq.(\ref{bararbitrary1}) we can get
\begin{eqnarray}
p_Ic_I^{(n)} &\approx &\kappa \gamma \hbar \left[ {\cal K}_I+\frac A{2\kappa
{\cal K}\hbar ^2}(p_1p_2p_3)^{\frac{1-w}2}\right]  \nonumber \\
&\approx &\kappa \gamma \hbar {\cal K}_I.  \label{bararbitrary7}
\end{eqnarray}
In the second step we use the condition that ${\cal K}_I\gg \frac A{2\kappa
{\cal K}\hbar ^2}(p_1p_2p_3)^{\frac{1-w}2}$. Substituting the above equation to
Eq.(\ref{bar1}) we have
\[
\frac 1{p_1}\frac{dp_1}{dt^{\prime }}\approx \kappa \hbar \cos (\bar{\mu}%
_1c_1)\mathfrak{S}_n(\bar{\mu}_1c_1)({\cal K}_2+{\cal K}_3).
\]
In the backward evolution, the $\bar{\mu}_1c_1$ gets more and more
significant, at some point $\cos (\bar{\mu}_1c_1)=0$ and the big bounce
occurs. Combining Eq.(\ref{higherholonomy}) and Eq.(\ref{bararbitrary7}) we
can get
\begin{eqnarray}
p_1^{3/2} &=&\sqrt{\triangle }\frac{\kappa \gamma \hbar {\cal K}_1}{%
\sum_{k=0}^n\frac{\left( 2k\right) !}{2^{2k}\left( k!\right) ^2(2k+1)}\left[
\sin \left( \bar{\mu}_1c_1\right) \right] ^{2k+1}}  \nonumber \\
&\geqslant &\sqrt{\triangle }\frac{\kappa \gamma \hbar {\cal K}_1}{%
\mathfrak{F}_n}.
\end{eqnarray}
In the second step the equality holds at the bouncing point, which is the same
as the massless scalar field. As a result the critical value of $p_I$ is
\begin{equation}
p_{I,crit}=\left[ \sqrt{\triangle }\frac{\kappa \gamma \hbar {\cal K}_1}{%
\mathfrak{F}_n}\right] ^{2/3}.  \label{bararbitrary11}
\end{equation}
We can also define the directional density $\rho _I$ as
\begin{equation}
\rho _I:=\frac{\kappa \hbar ^2{\cal K}_I^2}{3p_I^3},
\end{equation}
with the expression of $\rho _I$ we can say that the big bounces take place
whenever each of the directional density reaches the critical value
\begin{equation}
\rho _{I,crit}=\frac 13\mathfrak{F}_n^2(\kappa \gamma ^2\triangle )^{-1}\sim
\rho _{pl}.
\end{equation}

Plugging $p_{I,crit}$ to the condition ${\cal K}_I\gg\frac{A}{2\kappa{\cal K}%
\hbar^2}(p_1p_2p_3)^{\frac{1-w}{2}}$ it can be found that for case (i) the
constant $A$ has to satisfy
\begin{equation}
A\ll\mathfrak{F}_n^{1-w}\frac{|\kappa_I|\gamma^{w-1}}{|\kappa_1\kappa_2%
\kappa_3|^{(1-w)}}{\cal K}^{1+w}\kappa^w\hbar^{1+w}\triangle^{\frac{w-1}{2}}.\label{bararbitrary12}
\end{equation}

Now we consider case (ii)(the isotropized phase), where the matter sector dominates and the universe
is isotropized. In this case Eq.(\ref{bararbitrary4}) turns out to be
\begin{equation}
f(t)\approx\sqrt{\frac{A}{3\kappa\hbar^2}}(p_1p_2p_3)^{\frac{1-w}{4}}.
\label{bararbitrary8}
\end{equation}
Then Eq.(\ref{bararbitrary1}) becomes
\begin{equation}
p_Ic_I^{(n)}\approx\gamma\sqrt{\frac{\kappa A}{3}}(p_1p_2p_3)^{\frac{1-w}{4}%
},  \label{bararbitrary9}
\end{equation}
where we use the condition ${\cal K}_I\ll\frac{A}{2\kappa{\cal K}\hbar^2%
}(p_1p_2p_3)^{\frac{1-w}{2}}$. From Eq.(\ref{bar1}) we can get
\begin{equation}
\frac{1}{p_1}\frac{dp_1}{dt^{\prime}}\approx2\sqrt{\frac{\kappa A}{3}}\cos(%
\bar{\mu}_1c_1)\mathfrak{S}(\bar{\mu}_1c_1)(p_1p_2p_3)^{\frac{1-w}{4}}.
\end{equation}
We can see that once again when $\cos(\bar{\mu}_1c_1)=0$, the bounce occurs.
Applying Eq.(\ref{higherholonomy}) to Eq.(\ref{bararbitrary9}) we have
\begin{equation}
p_{1,crit}^{3/2}=\frac{\gamma}{\mathfrak{F}_n}\sqrt{\frac{\kappa A\triangle}{%
3}}(p_{1,crit}p_2p_3)^{\frac{1-w}{4}}.
\end{equation}
It can be seen that the critical value of $p_I$ is coupled with other
directions. If we assume in the isotropized case the bouncing points in different
directions are roughly at only slightly different moments, we can get the
approximated critical value for different $p_I$:
\begin{equation}
p_{crit}=\left[\frac{\kappa A\triangle\gamma^2}{3\mathfrak{F}_n^2}\right]^{%
\frac{2}{3+3w}}.  \label{bararbitrary10}
\end{equation}
This means that the critical density of the bouncing point is
\begin{equation}
\rho_{crit}=Ap^{-3(1+w)/2}_{crit}=3\mathfrak{F}_n^2(\kappa
\triangle\gamma^2)^{-1}\sim\rho_{pl}.
\end{equation}
And in this case the criterion is
\begin{equation}
A\gg\gamma^{w-1}{\cal K}^{1+w}\kappa_I^{\frac{1+w}{2}}\hbar^{1+w}\kappa^w%
\triangle^{\frac{w-1}{2}}\mathfrak{F}_n^{1-w}.\label{bararbitrary13}
\end{equation}

Finally, we turn to case (iii)(the transition phase). We can see that the order of magnitude of the right hand side of Eq.(\ref{bararbitrary12}) and Eq.(\ref{bararbitrary13}) is nearly the same. So in the transition phase, the criterion is also between the one in the two phases:
\begin{equation}
A\sim\gamma^{w-1}{\cal K}^{1+w}\kappa_I^{\frac{1+w}{2}}\hbar^{1+w}\kappa^w%
\triangle^{\frac{w-1}{2}}\mathfrak{F}_n^{1-w},
\end{equation}
and the bouncing points of $p_I$ is between the critical value of $p_I$
given in Eq.(\ref{bararbitrary11}) and Eq.(\ref{bararbitrary10}).

The mean scale factor $a(t)$ of the three cases with $w=1/3$ (radiation field)
in $\bar{\mu}$ scheme is depicted in Fig.\ref{fig01}(b),(c), and (d). In each
case the singularity is replaced by the big bounce regardless of the order
of holonomy corrections. As the matter contribution is more and more
dominant, the difference of the evolution with different holonomy orders is
more and more inconspicuous.

In \cite{Singh} it was shown that in the $\bar{\mu}$ scheme the critical value of energy density fails to have an upper bound. This can be seen as follows. In the traditional holonomy corrections, it was derived that the energy density has the form $
\rho_M=\frac{1}{\kappa\gamma^2\triangle}\big(\frac{\sqrt{p_1p_2}}{p_3}\sin(\bar{\mu}_1c_1)\sin(\bar{\mu}_2c_2)+ cyclic\ \ terms\big)
$.
As the prefactors such as $\frac{\sqrt{p_1p_2}}{p_3}$ are not bounded, $\rho_M$ are not bounded above. In this paper our analysis is consistent with this conclusion. Although the $p_I$ has the lower bound Eq.(\ref{bararbitrary11}), it has no upper bound. As a result the factor $\frac{\sqrt{p_1p_2}}{p_3}$ can diverge when $p_2\rightarrow\infty$ or $p_3\rightarrow\infty$. The numerical simulations in Fig.\ref{fig01} shows the solutions of the bounce, but the bounce is not generic.

\begin{widetext}
\begin{center}
\begin{figure}[!ht]
\includegraphics[clip,width=0.75\textwidth]{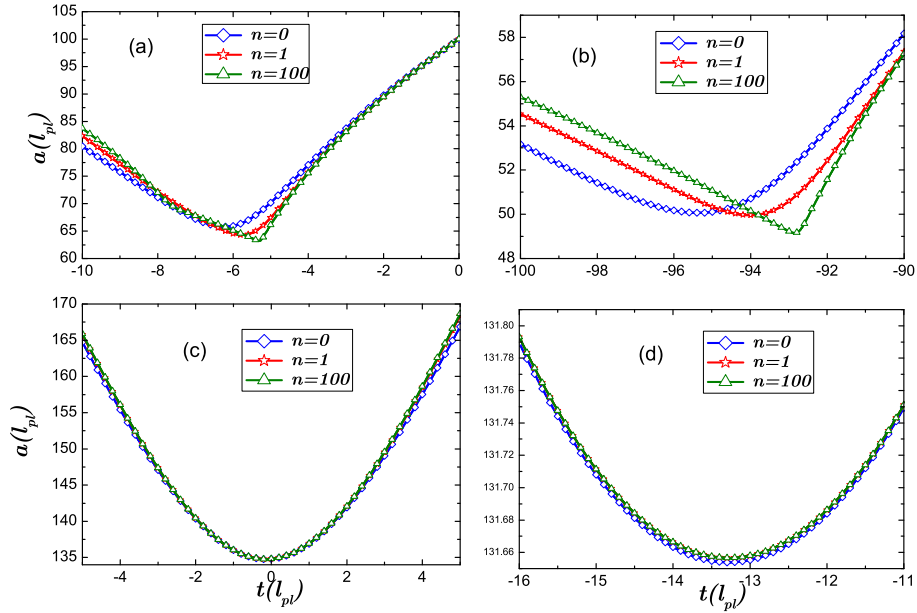}
\caption{Mean scale factor a(t) in $\bar{\mu}$ scheme corresponding to different orders of holonomy corrections. (a) for a massless scalar field with $\kappa_1=-1/4$, $\kappa_2=3/4$, $\kappa_3=1/2$, and $\kappa_\phi=1/\sqrt{8}$, $p_1(0)=p_2(0)=p_3(0)=10^4l_{pl}$, and $p_\phi=2\times 10^3\hbar\sqrt{\pi G}$. (b),(c), and (d) all for the radiation field with $w=1/3$, $\kappa_1=-2/7$, $\kappa_2=3/7$, $\kappa_3=6/7$, and $\mathcal{K}=1\times10^3$; (b) Kasner phase: $A=0.1\hbar l_{pl}^2$, $p_1(0)=3\times 10^4 l_{pl}$, $p_2(0)=2\times 10^4 l_{pl}$, and $p_3(0)=1\times 10^4 l_{pl}$. (c) Isotropized phase: $A=10^4\hbar l_{pl}^2$, $p_1(0)=9\times 10^4 l_{pl}$, $p_2(0)=6\times 10^4 l_{pl}$, and $p_3(0)=3\times 10^4 l_{pl}$. (d)Transition phase: $A=10^2\hbar l_{pl}^2$, $p_1(0)=3\times 10^4 l_{pl}$, $p_2(0)=2\times 10^4 l_{pl}$, and $p_3(0)=1\times 10^4 l_{pl}$.  }
\label{fig01}
\end{figure}
\end{center}
\end{widetext}

\section{\label{s5}Effective dynamics in $\bar{\mu}^{\prime}$ Scheme}

In this section we consider an alternative quantization scheme called $\bar{%
\mu}^{\prime }$ scheme. Previous works about $\bar{\mu}'$ scheme focus on a massless scalar field as it is difficult to get the analytical solution with arbitrary matter. Here we  extend the effective dynamics to arbitrary matter. The method used here is exactly the same as the $\bar{\mu}$ scheme and to avoid repetition we does not introduce it again in detail.  As it has no difference from the arbitrary matter case, we won't discuss the massless scalar field separately.

 For simplicity, we choose the lapse function $N=1/%
\sqrt{p_1p_2p_3}$ and introduce a new time variable
 $dt''=(p_1p_2p_3)^{1/2}dt$.  We also define a new variable
\begin{equation}
b_I^{(n)}=\bar{\mu}'_Ic_I^{(n)}.
\end{equation}
From the canonical relations we can get
\begin{eqnarray}
\left\{ c_I,b_J^{(n)}\right\}  &=&\kappa \gamma \frac{\bar{\mu}_J^{\prime }}{%
2p_I}\mathfrak{S}_n\left( \bar{\mu}_J^{\prime }c_J\right) \cos \left( \bar{%
\mu}_J^{\prime }c_J\right) c_J  \nonumber \\
&&for\ \ I=J,
\end{eqnarray}
\begin{eqnarray}
\left\{ c_I,b_J^{(n)}\right\}  &=&-\kappa \gamma \frac{\bar{\mu}_J^{\prime }%
}{2p_I}\mathfrak{S}_n\left( \bar{\mu}_J^{\prime }c_J\right) \cos \left( \bar{%
\mu}_J^{\prime }c_J\right) c_J,  \nonumber \\
&&for\ \ I\neq J
\end{eqnarray}
\begin{equation}
\left\{ p_I,b_J^{(n)}\right\} =-\kappa \gamma \bar{\mu}_J^{\prime }%
\mathfrak{S}_n\left( \bar{\mu}_J^{\prime }c_J\right) \cos \left( \bar{\mu}%
_J^{\prime }c_J\right) \delta _{IJ}.
\end{equation}
The Hamiltonian can be written as
\begin{equation}
\mathcal{H}_{\bar{\mu}'}=-\frac 1{\kappa \gamma ^2\triangle }\left[
b_1^{(n)}b_2^{(n)}+b_2^{(n)}b_3^{(n)}+b_3^{(n)}b_1^{(n)}\right] +\rho _M.\label{Hamiltonianconstraint}
\end{equation}
The equations of motion are
\begin{eqnarray}
\frac{dc_1}{dt^{\prime \prime }} &=&-\frac{\mathfrak{S}_n(\bar{\mu}%
_1^{\prime }c_1)\cos (\bar{\mu}_1^{\prime }c_1)\bar{\mu}_1^{\prime
}c_1\left[ b_2^{(n)}+b_3^{(n)}\right] }{2\gamma \triangle p_1}  \nonumber \\
&&+\frac{\mathfrak{S}_n(\bar{\mu}_2^{\prime }c_2)\cos (\bar{\mu}_2^{\prime
}c_2)\bar{\mu}_2^{\prime }c_2\left[ b_3^{(n)}+b_1^{(n)}\right] }{2\gamma
\triangle p_1}  \nonumber \\
&&+\frac{\mathfrak{S}_n(\bar{\mu}_3^{\prime }c_3)\cos (\bar{\mu}_3^{\prime
}c_3)\bar{\mu}_3^{\prime }c_3\left[ b_1^{(n)}+b_2^{(n)}\right] }{2\gamma
\triangle p_1}  \nonumber \\
&&+\kappa \gamma \frac{\partial \rho _M}{\partial p_1},  \label{barpry1}
\end{eqnarray}
\begin{equation}
\frac{dp_1}{dt^{\prime \prime }}=\frac{\mathfrak{S}_n(\bar{\mu}'_1c_1)\cos (%
\bar{\mu}'_1c_1)\bar{\mu}'_1(b_2^{(n)}+b_3^{(n)})}{\gamma \triangle }.  \label{barpry2}
\end{equation}
Combining Eq.(\ref{barpry1}) and Eq.(\ref{barpry2}) we can get
\begin{equation}
\frac d{dt^{\prime \prime }}\left( p_Ic_I-p_Jc_J\right) =0  \label{barpry3}
\end{equation}
In the above equation we used the assumption that the matter has zero anisotropy as in Sec.\ref{s2b}. As in the classical theory, $p_Ic_I-p_Jc_J$ is a constant. However, unlike
the classical case Eq.(\ref{classicalc}) is no longer satisfied. From Eq.(%
\ref{barpry2}) we can get the relation
\begin{equation}
\frac 1{p_1}\frac{dp_1}{dt^{\prime \prime }}=\frac 1{\gamma \sqrt{\triangle
p_1p_2p_3}}\mathfrak{S}_n(\bar{\mu}'_1c_1)\cos (\bar{\mu}'_1c_1)%
\left[ b_2^{(n)}+b_3^{(n)}\right] .
\end{equation}
The above equation has shown that as in the $\bar{\mu}$ case, when $\cos \left( \bar{\mu}%
_1^{\prime }c_1\right) =0$ the big bounce occurs. Now we investigate the
bouncing regime. From Eq.(\ref{barpry3}), we can assume the relation
\begin{equation}
p_Ic_I=\kappa \gamma \hbar \left[ {\cal K}_I+f(t)\right] .
\end{equation}
Note that in this equation the time-dependent part is the same for all of $%
p_Ic_I$, which is different from the $\bar{\mu}$ case. The energy density is
also assumed to be Eq.(\ref{rho}). Following the same derivation as in Sec.%
\ref{s4a} we can get Eq.(\ref{bararbitrary3}), Eq.(\ref{bararbitrary4})
and Eq.(\ref{bararbitrary5}). Here we also consider three cases separately.

(i) The Kasner phase:

In this phase Eq.(\ref{bararbitrary6}) holds and we have
\begin{equation}
p_Ic_I\approx \kappa \gamma \hbar {\cal K}_I.  \label{barpry4}
\end{equation}
At the bouncing point, $\cos(\bar{\mu}'_Ic_I)=0$, which corresponds to $\bar{\mu}'_Ic_I=\pi/2$. Applying this equation to Eq.(\ref{barpry4}) we can get the criterion of the bouncing point of $p_I$:
\begin{equation}
\sqrt{p_1p_2p_3}=\frac{2}{\pi}\kappa\gamma \hbar \sqrt{\triangle }{\cal K}_I.
\end{equation}
The critical energy density at this point is
\begin{equation}
\rho _{crit,I}=A\left[ \frac 2\pi \kappa \gamma \hbar \sqrt{\triangle }{\cal %
K}_I\right] ^{-(1+w)}.  \label{barpry5}
\end{equation}
As in the $\bar{\mu}$ case, the criterion for the Kasner case is also ${\cal %
K}_I\gg \frac A{{\cal K}\kappa \hbar ^2}\left( p_1p_2p_3\right) ^{\frac{1-w}2%
}$. Applying the critical value of $\sqrt{p_1p_2p_3}$ to this condition we
can get
\begin{equation}
A\ll (2/\pi )^{w-1}\kappa ^w\gamma ^{w-1}{\cal K}^{w+1}\kappa _I^w\hbar
^{w+1}\triangle ^{\frac{w-1}2}.
\end{equation}
Applying it to Eq.(\ref{barpry5}) we have
\begin{equation}
\rho _{M,crit}\ll \left( \frac \pi 2\right) ^2\kappa _I^{-1}\rho _{pl}.
\label{barpry6}
\end{equation}

(ii) The isotropized phase:

In this case, Eq.(\ref{bararbitrary8}) holds and we have
\begin{equation}
p_Ic_I=\gamma \sqrt{\frac{\kappa A}3}\left( p_1p_2p_3\right) ^{\frac{1-w}4}.
\end{equation}
Applying the bouncing condition $\cos(\bar{\mu}'_Ic_I)=0$ as in the Kasner phase we can get
\begin{equation}
\sqrt{p_1p_2p_3}=\left( \frac \pi 2\right) ^{\frac 2{1+w}}\left( \frac{%
\gamma ^2\kappa A\triangle }3\right) ^{\frac 1{1+w}}.
\end{equation}
One can see that for different $p_I$ the condition at the bouncing point is
the same, which means that all $p_I$ at three directions bounce at the same
time. The criterion for this case is
\begin{equation}
A\gg (2/\pi )^{w-1}\kappa ^w\gamma ^{w-1}{\cal K}^{w+1}\kappa
_I^{(1+w)/2}\hbar ^{w+1}\triangle ^{\frac{w-1}2}.
\end{equation}
The critical value of energy density is
\begin{equation}
\rho _{crit}=\left( \frac \pi 2\right) ^2(\gamma ^2\frac{\kappa \triangle }3%
)^{-1}=3\left( \frac \pi 2\right) ^2\rho _{Pl}\sim 10\rho _{pl}.
\label{barpry7}
\end{equation}

(iii) The transition phase:

Here $A\sim (2/\pi )^{w-1}\kappa ^w\gamma ^{w-1}{\cal K}^{w+1}\kappa
_I^{(1+w)/2}\hbar ^{w+1}\triangle ^{\left( w-1\right) /2}$ and the critical
value of energy density is between the one got in Eq.(\ref{barpry6}) and Eq.(%
\ref{barpry7}).

The numerical solutions of mean scale factor $a(t)$ is depicted in Fig.\ref
{fig2}. The singularity is never approached and the big bounce of $a(t)$
occurs at any case. As the order of holonomy corrections increases, the big
bounce takes place more and more abruptly.

In previous investigations about the $\bar{\mu}'$ scheme, it was shown that there is a generic strong singularity resolution\cite{Singh,Singh01,Singh02,Singh04}. Here we can extend the conclusion to the case of higher order holonomy corrections. The vanishing of the Hamiltonian constraint Eq.(\ref{Hamiltonianconstraint}) leads to the expression of the energy density:
\begin{equation}
\rho_M=\frac 1{\kappa \gamma ^2\triangle }\left[
b_1^{(n)}b_2^{(n)}+b_2^{(n)}b_3^{(n)}+b_3^{(n)}b_1^{(n)}\right]
\end{equation}
Since the $b_I^{(n)}$ in the parenthesis are all bounded functions, the maximum value of the energy density is
\begin{equation}
\rho_{max}=\frac{3\mathfrak{F}_n^2}{\kappa\gamma^2\triangle}
\end{equation}
where the $\mathfrak{F}_n$ is defined in Eq.(\ref{Fn}). When $n\rightarrow\infty$, the $\rho_{max}\rightarrow \frac{3\pi^2}{4\kappa\gamma^2\triangle}$. For different $n$ the specific value of $\rho_{max}$ is different but still finite. The $\rho_{max}$ is the largest value of energy density in theory, but the real evolution of the universe may not attain it.

\begin{widetext}
\begin{center}
\begin{figure}[!ht]
\includegraphics[clip,width=0.8\textwidth]{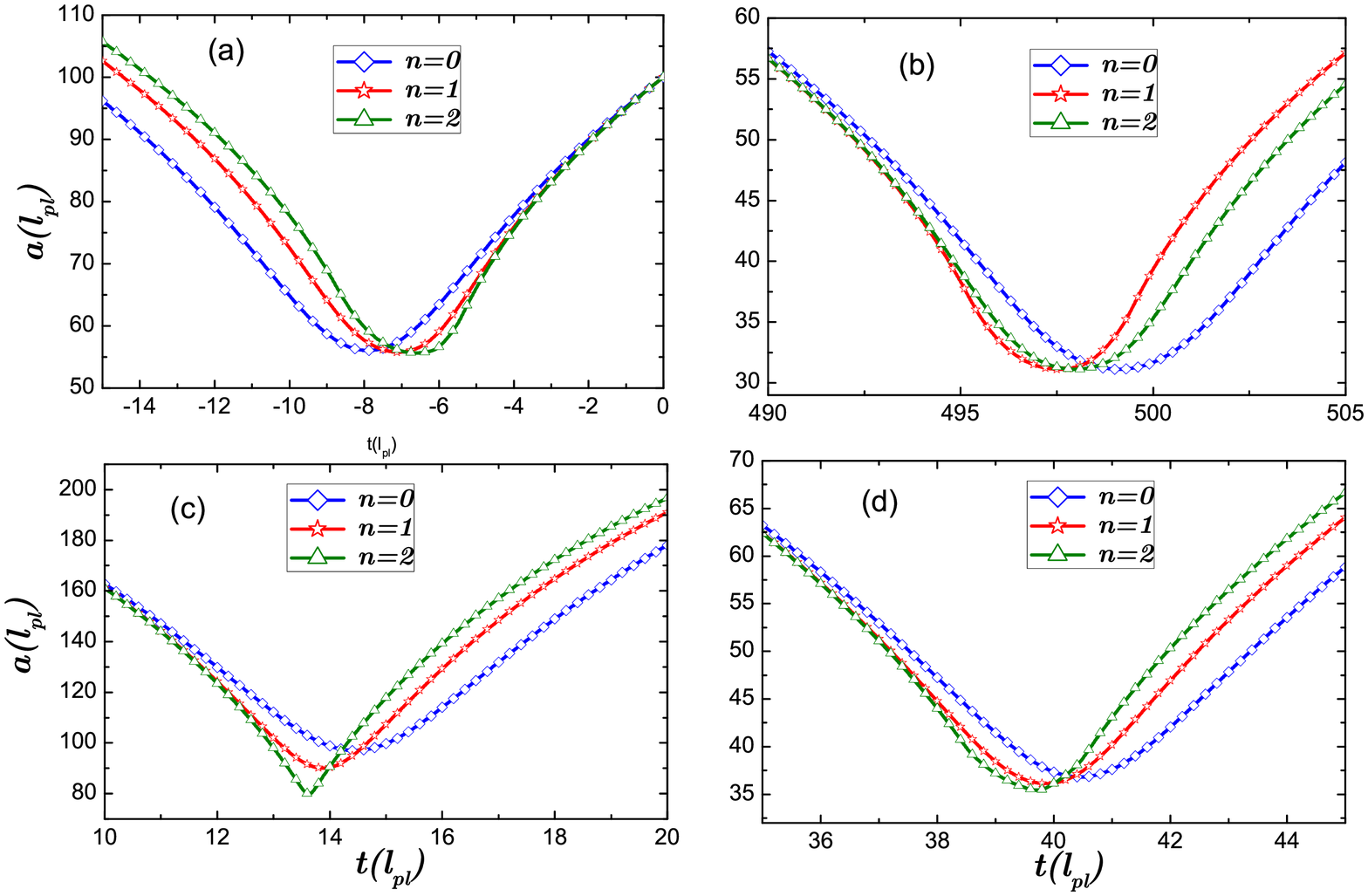}
\caption{Mean scale factor a(t) in $\bar{\mu}^{\prime}$ scheme corresponding to different orders of holonomy corrections. (a) for a massless scalar field case with $\kappa_1=-1/4$, $\kappa_2=3/4$, $\kappa_3=1/2$, $\kappa_\phi=1/\sqrt{8}$, $p_1(0)=p_2(0)=p_3(0)10^4l_{pl}$, and $p_\phi=2\times 10^3\hbar\sqrt{\pi G}$. (b), (c), and (d) all for the radiation field with $w=1/3$,$\kappa_1=-2/7$,$\kappa_2=3/7$,$\kappa_3=6/7$;$\mathcal{K}=1\times10^3$; (b)Kasner phase. With $A=0.1\hbar l_{pl}^2$; $p_1(0)=3\times 10^4 l_{pl}$, $p_2(0)=2\times 10^4 l_{pl}$, $p_3(0)=1\times 10^4 l_{pl}$. (c)Isotropized phase. With$A=10^4\hbar l_{pl}^2$; $p_1(0)=9\times 10^4 l_{pl}$, $p_2(0)=6\times 10^4 l_{pl}$, $p_3(0)=3\times 10^4 l_{pl}$.(d)Transition phase. With $A=10^2\hbar l_{pl}^2$; and $p_1(0)=3\times 10^4 l_{pl}$, $p_2(0)=2\times 10^4 l_{pl}$, $p_3(0)=1\times 10^4 l_{pl}$. }
\label{fig2}
\end{figure}
\end{center}
\end{widetext}

In \cite{Ashtekar},\cite{Chiou3} and \cite{Chiou4}, it was shown that in the $\bar{\mu}'$ scheme there is no directional dependence on the critical energy density, which seems contrary to our conclusions in this section, at least at the Kasner phase. This problem comes from the confusing definition of the critical energy density. If we define it to be the critical value at the bouncing point of the whole volume(corresponding to the bouncing point of $a$ defined in Eq.(\ref{a})), it is indeed directional independent. On the other hand, if we define the critical energy density to be the value at the bouncing point of directional Hubble parameter $H_I$, it is directional dependent. In general, in the Bianchi I model, there are three bounces of different directions, which occur at  distinct times. The criterion of bounce of a definite directional $p_I$ is Eq.(\ref{barpry5}), but this is not the bouncing point of the whole volume. As a simplest example, supposing $\rho_{crit,1}<\rho_{crit,2}<\rho_{crit,3}$, when the whole energy density reaches $\rho_{crit,1}$, the $p_1$ bounces, but $p_2$ and $p_3$ are still contracting, which is still leading to the contraction of the whole volume. Similarly, when $\rho=\rho_{crit,2}$, the $p_2$ bounces, but the whole volume is still contracting with $p_3$. Only at the bouncing point of $p_3$ does the whole volume bounce, namely in this case the critical energy density of the whole volume is just $\rho_{crit,3}$. Eq.(\ref{barpry5}) shows that the magnitude of the critical energy densities of different $p_I$ are dominated by $\mathcal{K}_I$, which are the integration constants that only relate to the initial conditions.

 In \cite{Chiou4}, the critical energy densities of three directions are nearly the same, but the result is under the isotropic approximation which corresponds to our 'isotropized phase'. In that paper it is also approved that when the solution is far from isotropic and the approximation is violated, the matter density is still the indication of the bounce occurrence, with the critical value quite different from each other.  Fig.\ref{fig2} actually depicts the bounce of the whole volume.

In fact, the case that the whole volume bounces at the point $\rho_{bounce}=\rho_{crit,3}$ as described above is the simplest situation. Comparing the value of $\rho_{bounce}$ to the three values of $\rho_{crit,I}$ can lead to an insight about an interesting phenomena called the 'Kasner transition' as investigated in  \cite{Singh02}. Generally speaking, the Kasner transition is to describe the transition of the geometry structure of space before and after the big bounce of the whole volume. For example, in the pre-bounce phase when all the three scale factors $a_I$ contracting or in the post-bounce phase all three $a_I$ expanding, it's called the point like structure, as the evolution forward in time in the pre-bounce phase or backward in the post-bounce phase will structure a point singularity in classical dynamics. Another structure is called cigar like when $a_1$ expanding but $a_2$ and $a_3$ contracting in the pre-bounce phase or $a_1$ contracting but $a_2$ and $a_3$ expanding in the post-bounce phase as the physical evolution forward in time in the pre-bounce or backward in the post-bounce phase will lead to an infinite cigar like sigularity in classical dynamics.

With the analytical investigations provided in this section, the Kasner transition described in \cite{Singh02} can be interpreted as follows. Supposing $\rho_{crit,1}<\rho_{crit,2}<\rho_{crit,3}$ in the point like structure of pre-bounce phase, if the bouncing point of the whole volume is at $\rho_{bounce}=\rho_{crit,3}$ as described above, after the bounce all three $a_I$ expanding, which is also the point like structure. This is the so-called point-point transition. If in the pre-bounce phase the structure is  cigar like, one can verify that the bouncing point at $\rho_{bounce}=\rho_{crit,3}$ can also lead to the cigar-cigar transition.

 The same as before, supposing $\rho_{crit,1}<\rho_{crit,2}<\rho_{crit,3}$ in the point like structure of pre-bounce phase, but this time the bouncing point is at $\rho_{crit,1}<\rho_{bounce}<\rho_{crit,2}$. After the bounce, the $p_1$ increases but $p_2$ and $p_3$ still decrease. After the bounce, the scale factor $a_2=\frac{\sqrt{p_1p_2p_3}}{p_2}$ and $a_3=\frac{\sqrt{p_1p_2p_3}}{p_3}$ expanding as the denominator decrease but the molecular increase. The only contracting scale factor is $a_1$ as $a_1\propto\frac{p_2}{a_3}$, which leads to the cigar like structure. This is the so-called point-cigar transition, which means that the spatial structures before and after the bounce are not the same. In this case, the $\rho_{crit,2}$ and $\rho_{crit,3}$ are never attained as after the bounce the energy desity
will decrease. But this does not mean the physical evolution would lead to a piece like structure. In \cite{Singh03} it was shown that in the cigar like structure of post-bounce phase the inflation model can lead the contracting direction to undergo a turn around in the forward evolution. But this procedure is not due to quantum bounce.

\section{\label{s6}Anisotropy}

In Bianchi I model the anisotropy is included. Here we use the shear parameter $%
\Sigma$ to describe the anisotropy.

In classical case, the anisotropy is discussed in Sec. {\ref{s2b}}. We can see
from Eq.(\ref{clshear}) that in this case the shear parameter is a constant,
which does not vary when the universe evolves. However, Eq.(\ref{shearscalar}%
) shows that when the singularity is reached, the shear scalar blows up as $%
a $ approaches $0$.

In the effective dynamics with higher order holonomy corrections, the shear
parameter $\Sigma $ is
\begin{equation}
\Sigma ^2 =\frac{a^6}{18}\left[ (H_1-H_2)^2+(H_2-H_3)^2+(H_3-H_1)^2\right]\label{shear}
\end{equation}
From Eq.(\ref{p}) we have
\begin{equation}
a_1=\frac{1}{l_1}\left(\frac{p_2p_3}{p_1}\right)^{\frac{1}{2}}
\end{equation}
and its cyclic permutations. As a result
\begin{equation}
H_1=\frac{\dot{a_1}}{a_1}=-\frac{\dot{p}_1}{2p_1}+\frac{\dot{p}_2}{2p_2}+\frac{\dot{p}_3}{2p_3}\label{H}
\end{equation}
Here the $\dot{a}$ is the derivative of $a$ with cosmic time $t$ which have the lapse function $N=1$.
From the definition of the Hamiltonian Eq.(\ref{Hamiltonian}) we can see when $N=1$ we have
\begin{eqnarray}
{\cal H}_{eff} &=&-\frac 1{\kappa \gamma ^2\sqrt{p_1p_2p_3}}\left[
c_1^{(n)}p_1c_2^{(n)}p_2+c_2^{(n)}p_2c_3^{(n)}p_3\right.   \nonumber \\
&&\left. +c_3^{(n)}p_3c_1^{(n)}p_1\right] +\sqrt{p_1p_2p_3}\rho_M.
\end{eqnarray} As a result
\begin{eqnarray}
\dot{p_1}&=&-\kappa\gamma\frac{\partial\mathcal{H}_{eff}}{\partial c_1}\nonumber\\
&=&\frac{1}{\sqrt{p_1p_2p_3}}\frac{1}{\gamma}\cos(\bar{\mu}_1c_1)\mathfrak{S}(\bar{\mu}_1c_1)p_1\left(c_2^{(n)}p_2+c_3^{(n)}p_3\right)\nonumber\\
~\label{dotp}
\end{eqnarray}
Applying Eq.(\ref{dotp}) to Eq.(\ref{H}) and then substitute the result to Eq.(\ref{shear}) we can get the following result:
\begin{eqnarray}
\Sigma ^2
&=&\frac 1{6\gamma ^2}\left\{ \left[ \cos (\bar{\mu}_2c_2)\mathfrak{S}_n(%
\bar{\mu}_2c_2)\left( p_1c_1^{(n)}+p_3c_3^{(n)}\right) \right. \right.
\nonumber \\
&&\left. -\cos (\bar{\mu}_1c_1)\mathfrak{S}_n(\bar{\mu}_1c_1)\left(
p_2c_2^{(n)}+p_3c_3^{(n)}\right) \right] ^2  \nonumber \\
&&+\left[ \cos (\bar{\mu}_3c_3)\mathfrak{S}_n(\bar{\mu}_3c_3)\left(
p_1c_1^{(n)}+p_2c_2^{(n)}\right) \right.   \nonumber \\
&&\left. -\cos (\bar{\mu}_2c_2)\mathfrak{S}_n(\bar{\mu}_2c_2)\left(
p_3c_3^{(n)}+p_1c_1^{(n)}\right) \right] ^2  \nonumber \\
&&+\left[ \cos (\bar{\mu}_1c_1)\mathfrak{S}_n(\bar{\mu}_1c_1)\left(
p_2c_2^{(n)}+p_3c_3^{(n)}\right) \right.   \nonumber \\
&&\left. \left. -\cos (\bar{\mu}_3c_3)\mathfrak{S}_n(\bar{\mu}_3c_3)\left(
p_1c_1^{(n)}+p_2c_2^{(n)}\right) \right] ^2\right\} . \nonumber\\
~ \label{qtshear}
\end{eqnarray}
Contrary to the classical case, the shear is not a
constant because of the holonomy corrections. Besides that, different orders
of holonomy corrections correspond to different values of shear. In the
classical regime, $\bar{\mu}_Ic_I\rightarrow 0$, which means $\cos (\bar{\mu}%
_Ic_I)=1$, $\mathfrak{S}_n(\bar{\mu}_Ic_I)=1$ and $c_I^{(n)}\rightarrow c_I$, the shear parameter turns out to be
\begin{eqnarray}
\Sigma ^2 &=&\frac 1{6\gamma ^2}\left[ \left( p_1c_1-p_2c_2\right) ^2+\left(
p_2c_2-p_3c_3\right) ^2\right.   \nonumber \\
&&\left. +\left( p_3c_3-p_1c_1\right) ^2\right] ,  \label{clformshear}
\end{eqnarray}
which goes back to the classical form.

\subsection{\label{s6a}Anisotropy in $\bar{\mu}$ scheme}

In $\bar{\mu}$ scheme with a massless scalar field, when we consider the
pre-bounce classical regime, the shear parameter turns out to be
\begin{eqnarray}
\Sigma ^2 &=&\frac{\kappa ^2\hbar ^2}6\left[ \left( {\cal K}_1-{\cal K}%
_2\right) ^2+\left( {\cal K}_2-{\cal K}_3\right) ^2+\left( {\cal K}_3-{\cal K%
}_1\right) ^2\right]   \nonumber \\
&&~  \label{323}
\end{eqnarray}
which is a constant. At the bouncing regime, although the shear parameter
changes its value, the $p_Ic_I^{(n)}$ keeps its constant value through the
bounce. After the bounce occurs and the classical behavior is recovered, the
constant $p_Ic_I^{(n)}$ is the same as the pre-bounce value. So we can
conclude that the shear does not change its value after the bounce. The
shear for massless scalar field is depicted in Fig.\ref{fig3}(a). One can
see the shear parameter in the pre-bounce and post-bounce regime is the
same. As $n$ increases, the shear parameter also changes more abruptly.

In the Kasner phase, the shear parameter is also given by Eq.(\ref{qtshear}). In the
pre-bounce classical case, the shear is in the form of Eq.(\ref{323}).
During the bouncing regime, although $\Sigma$ varies as $\bar{\mu}_Ic_I$ gets
more and more significant, the $p_Ic_I^{(n)}\approx{\cal K}_I$, which is
nearly a constant. When the evolution approaches the classical regime again,
its value comes back to Eq.(\ref{323}). As a result, we can conclude that in
the Kasner phase, we have $\Sigma^2(post\ \ bounce)\approx\Sigma^2(pre\ \
bounce) $.

Now we discuss the shear in the isotropized case. In this case, the
expression is also Eq.(\ref{qtshear}). However, we can see from Eq.(\ref
{bararbitrary9}) that the $p_Ic_I^{(n)}\approx \gamma \sqrt{\frac{\kappa A}3}%
\left( p_1p_2p_3\right) ^{\left( 1-w\right) /4}$, which can not keep its
value as a constant as $p_I$ varies. So in general the shear parameter $%
\Sigma $ is not the same in the post-bounce and pre-bounce regimes as the
information of anisotropy is smeared.

The shear parameter with arbitrary matter in $\bar{\mu}$ scheme is depicted
in Fig.\ref{fig3}(b). It shows that when the order of holonomy corrections
increases, the variation of the shear value between post-bounce and
pre-bounce regimes gets more and more smaller. When $n\rightarrow \infty $,
we have $\Sigma ^2(post\ \ bounce)=\Sigma ^2(\left( pre\ \ bounce\right) $.

We have to note the case where the order of holonomy corrections $n=\infty $%
. In this case, $c_I^{(n)}=c_I$, $\mathfrak{S}_n(\bar{\mu}_Ic_I)=1/\left|
\cos \left( \bar{\mu _Ic_I}\right) \right| $, one may have the idea that the
quantum form of shear parameter $\Sigma$ turns out to be the classical form ,which may mean
that the shear parameter $\Sigma $ is a constant not only in the classical
regime but also in the bouncing regime. However, Fig.\ref{fig3}(a) shows that
it's not the case. The reason is that different $p_I$ bounce at different
moments, as a result different $\cos (\bar{\mu}_Ic_I)$ flip signs at
different time. From Eq.(\ref{qtshear}) one can see that if $\cos({\mu}_Ic_I)$ have different signs,
the form of $\Sigma$ can't reduce to the classical form.  So even in the case where $n=\infty $ the shear parameter
does vary its value at the bouncing regime, although the change is abrupt.

\subsection{\label{s6b}Anisotropy in $\bar{\mu}^{\prime}$ Scheme}

At last we discuss the shear in the $\bar{\mu}^{\prime}$ scheme. The
expression of the shear $\Sigma^2$ is also Eq.(\ref{qtshear}). In the
classical regime, it reduces to Eq.(\ref{clformshear}). From Eq.(\ref
{barpry3}) we know that $p_Ic_I-p_Jc_J$ is a constant all around the
evolution process, including the bouncing regime. As a result, the shear is
the same in the pre-bounce classical regime and the post-bounce classical
regime in any case.

The shear parameter with a massless scalar field and arbitrary matter is
depicted in Fig.\ref{fig3}(c) and Fig.\ref{fig3}(d). We can see the
value of pre-bounce regime is the same as post-bounce regime, regardless of
the order of holonomy corrections. When $n\rightarrow\infty$, the shear does
change its value in the bouncing regime as the $\bar{\mu}$ scheme.

\begin{widetext}
\begin{center}
\begin{figure}[!ht]
\includegraphics[clip,width=0.8\textwidth]{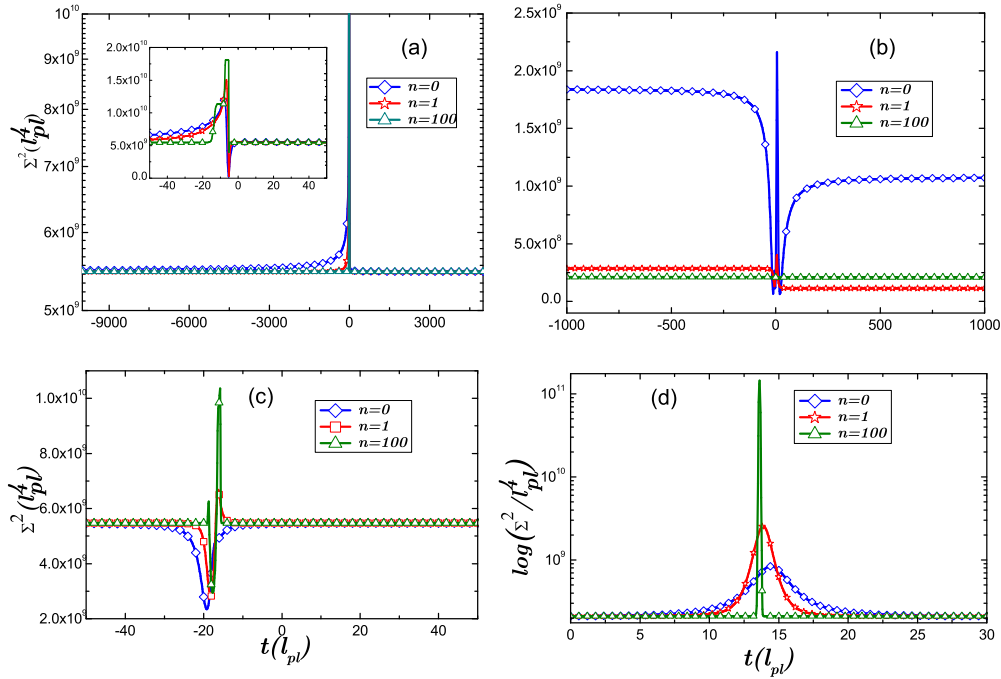}
\caption{Shear term in $\bar{\mu}$ and $\bar{\mu}'$ schemes for different orders of holonomy corrections. (a)Massless scalar field in $\bar{\mu}$ scheme. (b)Arbitrary matter in $\bar{\mu}$ scheme. (c)Massless scalar field in $\bar{\mu}'$ scheme. (c)Arbitrary matter in $\bar{\mu}'$ scheme.}.  \label{fig3}
\end{figure}
\end{center}
\end{widetext}

\section{\label{s7}Summary and conclusions}

In this paper we investigate the Bianchi I model in LQC with higher order
holonomy corrections in the heuristic effective level.

The focus of this paper is to construct the effective dynamics with higher order holonomy corrections
in the form of $\bar{\mu}'$ scheme. In this scheme, due to the mathematical complexity most of previous investigations have focused on the numerical simulations. Here we try to study this scheme with analytical method. To compare with the $\bar{\mu}'$ scheme and the earlier works in Bianchi I model we also investigate the $\bar{\mu}$ scheme, respectively.
 The $\bar{\mu}$ scheme has been proven to be a problematic scheme as it's fiducial cell dependent and fails to solve the strong singularity. Here we find that in this scheme with higher order holonomy corrections, we can really find out the solutions of the big bounce, but even so the energy density has no generic upper bound, which means in the $\bar{\mu}$ scheme the strong singularity is not resolved.  Contrary to this, in the $\bar{\mu}'$ scheme the strong singularity is never reached and the big bounce is robust
regardless of the order of the
holonomy corrections, even when the order is $\infty$, which is consistent with the recent investigations\cite{Singh,Singh01,Singh02,Singh04}.

 Besides that, we also find out some detailed difference in the two schemes. In the case of $\bar{\mu}$ scheme with a massless scalar field and with an arbitrary matter field in the
Kasner phase, different $p_I$ bounce at different time, and the critical
values of $p_I$ can be found out. We use the directional density $\rho_I$ to
describe the bounces. When a directional density $\rho_I$ reaches its
critical value, the $p_I$ bounces. In $\bar{\mu}$ scheme with isotropized
phase, the three directions of $p_I$ bounce roughly at the same time. When
the energy density reaches the critical value the bounces occur at all three
directions. When the higher order holonomy corrections are considered,
  the critical value is given by the one in the conventional case times a numerical factor $\mathfrak{F}$. For different orders of holonomy
corrections the numerical factor is also different.

Different from the $\bar{\mu}$
scheme, we find that in the $\bar{\mu}'$ scheme the critical values of $p_I$ are not clear
but we can find out the critical energy densities, and if the energy density
reaches one of the critical values a bounce happens. In the Kasner phase different
directions correspond to different critical densities while for the
isotropized phase all three directions have the same critical density and
the bounces happen at the same time. We also find that in the Kasner case
the critical energy densities are far less than the Planck density. In the
isotropized case the value approaches the Planck density, which is the same
as the result in \cite{Chiou4}. Besides that, it is shown that the critical
densities are not depend on the order of holonomy corrections, which is
different from the $\bar{\mu}$ scheme.

In the $\bar{\mu}'$ scheme we also find out an interesting phenomena that given the criterion of the three directions of $p_I$(Eq.(\ref{barpry5})), weather the physical evolution can attain all of them dominates the spatial structure after the big bounce. This leads to the so-called 'Kasner transition' as investigated in \cite{Singh02}. We find that when the largest value of the three critical energy density can be attained, the point-point transition or cigar-cigar transition comes into being. When just the minimum of the three critical energy densities can be attained, the point-cigar transition can be formed.  In \cite{Singh02}, with the numerical simulations it was found that the Kasner transition follows selection rules which are determined by the initial relative strenth of the matter and anisotropy. In this paper we just attempt to interpret it analytically and this is not the focus of this paper. Our following works would be  investigating this issue systematically in the analytical method and find out the exact condition for every Kasner transition case and compare with the numerical simulations.

We use the shear parameter $\Sigma$ to describe the anisotropy. In the
classical case, the shear parameter is a constant. In the semiclassical
regime the shear parameter is not a constant, which is a kind of quantum
effects. In the $\bar{\mu}$ scheme under the condition that the order of holonomy corrections  is finite,
generically the shear parameter can not
keep its constant value before and after the bounce unless at the massless scalar field case.  Contrary to the $\bar{\mu}$ scheme, in the $\bar{\mu}^{\prime}$
scheme the shear parameter can hold its value in
 classical regimes before and after the bounce  in all cases. When we consider the case where $n=\infty$, the shear parameter does change its value in the bouncing regime, which indicates that the variation of the shear parameter is a real quantum effect
which does not depend on the artificial choice of the order of the holonomy
corrections.

It's interesting when the order of holonomy corrections approaches $\infty$.
In this case, the expression of the moving equations turn out to be the
classical form, even in the bouncing regime, but the dynamics is not the same as the classical case. The reason is shown in Sec.\ref{s3} that the power series $c_I^{(\infty)}$ gives back to $c_I$ only when $-\pi/2\leqslant\bar{\mu}_Ic_I\leqslant\pi/2$, and out of this regime $c_I^{(n)}$ does not equal to $c_I$. When $n\rightarrow\infty$, in the contracting phase before bounce, the directional Hubble
parameter $H_I$ never gets to be $0$,  at the bouncing point the $H_I$
changes its sign abruptly and the $p_I$ transforms from the contracting
phase to the expanding phase. In other words, the Hubble parameter is
discontinuous at this point. This is the picture of motion in the
semiclassical level, which is different from previous investigations. Surely
we know that at the bouncing point  the semiclassical dynamics
is only an approximation, this picture is closer to the quantum picture than
before.

 It is worth noting that we can have $c_I^{(n)}\rightarrow c_I$  no matter when $\bar{\mu}_I\rightarrow 0$ or $n\rightarrow \infty$, but the two cases are not the same. The first reduces the dynamics to the classical case while the second one has a kink at the bounce point $\bar{\mu}_Ic_I=\pi/2$.

The method of higher order holonomy corrections is a promising approach. In
fact, some other problems may be solved when the order of holonomy
corrections approaches $\infty$, i.e. the anomaly free problem in the
perturbation theory of LQC \cite{Bojowald3,Bojowald4,Y.Ling,Mielczarek}.
The previous solution is to add counter terms to the Hamiltonian to eliminate
the anomaly terms, but when we consider the infinite orders of holonomy
corrections the anomaly terms disappear spontaneously. Whether there are
other qualitative changes when all orders of holonomy corrections are
considered still requires future investigations.

Now it seems that at the effective level the order of the holonomy
corrections  can vary from $0$ to $\infty$ and the
definite value of $n$ is not clear. In \cite{Chiou6} the quantum approach is constructed
in homogeneous and isotropic cosmology, which shows that the higher order holonomy corrections is related to the $j$ ambigurity. In Bianchi I model, what is the definite meaning of
higher order holonomy corrections has not been thoroughly studied and we can only speculate  from the isotropic case.  In addition, whether $n$ has to approach $\infty$ are still
questionable.   The construction of the rigorous quantum theory with higher order
holonomy corrections in Bianchi I model  is
necessary to answer these questions, which is still an open issue. As well as this, even in the isotropic case the higher order holonomy corrections is still not conclusive. The investigations about this may shed light to the issue of $j$ ambiguity and the link between LQC and spin-foams.

\acknowledgments This work was supported by the National Natural Science Foundation of China under Grant Nos. 11175019 and 11235003.


\begin{thebibliography}{99}
\bibitem{LQG}T. Thiemann, Lect. Notes Phys. {\bf 631}, 41 (2003); A. Ashtekar and J. Lewandowski, Class. Quan. Grav. {\bf 21}, R53 (2004); A. Ashtekar, arXiv:1201.4598; C. Rovelli, {\it Quantum Gravity}, Cambridge Monographs on Mathematical Physics (Cambridge University Press, Cambridge, England, 2004).

\bibitem{LQC}M. Bojowald, Living Rev. Relativity {\bf 8}, 11 (2005) (http://relativity.livingreviews.org/Articles/lrr-2005-11/); M.
Bojowald, Phys. Rev. Lett. {\bf 86}, 5227 (2001); M. Bojowald, G. Date, and
K. Vandersloot, Class. Quan. Grav. {\bf 21}, 1253 (2004); P. Singh and A.
Toporensky, Phys. Rev. D {\bf 69}, 104008 (2004); A. Ashtekar and P. Sing, Class. Quan. Grav.
{\bf28}, 213001 (2011).
\bibitem{LQC1}A. Ashtekar, T. Pawlowski, and P. Singh, Phys. Rev.
Lett. {\bf96}, 141301 (2006); A. Corichi, P. Singh, Phys. Rev. D {\bf78}, 024034 (2008).
\bibitem{LQC2}A. Ashtekar, M. Bojowald and J. Lewandowski, Adv.
Theor. Math. Phys. {\bf 7}, 233 (2003).

\bibitem{Chiou1}D. W. Chiou, Phys. Rev. D {\bf 75}, 024029 (2007).
\bibitem{Ashtekar}A. Ashtekar and E. Wilson-Ewing, Phys. Rev. D {\bf79} 083535 (2009).
\bibitem{Chiou2}D. W. Chiou,  gr-qc/0703010v2.
\bibitem{Chiou3}D. W. Chiou and K. Vandersloot, Phys. Rev. D {\bf 76},084015 (2007).

\bibitem{Chiou4}D. W. Chiou, Phys. Rev. D {\bf 76},124037 (2007).


\bibitem{Singh}A. Corichi and P.Singh, Phys. Rev.D {\bf 80}, 044024 (2009).
\bibitem{Singh01}B.Gupt and P.Singh, Phys. Rev.D {\bf 85}044011 (2012).
\bibitem{Singh02}B.Gupt and P.Singh, 	Phys. Rev. D {\bf 86}, 024034 (2012).
\bibitem{Singh03}B.Gupt and P.Singh, Class. Quantum Grav. 30 145013 (2013).
\bibitem{Singh04}P.Singh, Phys. Rev. D {\bf 85}, 104011 (2012).
\bibitem{Lukasz}Lukasz Szulc, Phys. Rev. D {\bf78},064035 (2008)

\bibitem{Singh1}P. Singh and K. Vandersloot, Phys. Rev. D {\bf 72}, 084004 (2005).
\bibitem{Singh2}P. Singh, K. Vandersloot and G. V. Vereshchagin, Phys. Rev. D {\bf 75}, 023523 (2007).
\bibitem{Chiou5}D. W. Chiou and Li-fang Li,  Phys. Rev. D {\bf 79},063510 (2009).
\bibitem{Chiou6}D. W. Chiou and Li-fang Li,  Phys. Rev. D {\bf 80},43512 (2009).

\bibitem{Ponzano}G. Ponzano and T. Regge,  Spectroscopy and
Group Theoretical Methods in Physics, Amsterdam: North-Holland (1968).
\bibitem{Freidel}L. Freidel and D. Louapre, Nucl. Phys. B 662, 279
(2003)
\bibitem{Baez}J. C. Baez,  Lect.
Notes Phys. 543, 25 (2000)
\bibitem{Noui}K. Noui and P. Roche,  Class.
Quant. Grav. 20, 3175 (2003)

\bibitem{Vandersloot}K. Vandersloot, Phys. Rev. D {\bf71},103506 (2005).
\bibitem{Perez}A. Perez, Phys. Rev. D {\bf73},044007 (2006).

\bibitem{BarberoImmirzi} K. A. Meissner, Class.Quant.Grav. {\bf21}, 5245 (2004).


\bibitem{Bojowald3}  M. Bojowald, G. M. Hossain and M. Kagan and S.
Shankaranarayanan, Phys. Rev. D {\bf 78}, 063547 (2008).
\bibitem{Bojowald4}M. Bojowald, G. M. Hossain, M. Kagan and S. Shankaranarayanan, Phys. Rev. D {\bf 79}  043505 (2009).
\bibitem{Y.Ling}J. P. Wu and Y. Ling, JCAP {\bf 1005},026.
\bibitem{Mielczarek} T. Cailleteau, J. Mielczarek, A. Barrau and J. Grain, Class. Quantum Grav. {\bf 29}, 095010 (2012).


\end{thebibliography}
\end{document}